\documentclass[a4paper,fleqn,usenatbib]{mnras}

\usepackage{newtxtext,newtxmath}

\usepackage[T1]{fontenc}
\usepackage{ae,aecompl}

\usepackage[dvipsnames]{xcolor}
\usepackage{float}
\usepackage{graphicx}	
\usepackage{amsmath,siunitx}	
\usepackage{mathtools}
\usepackage{tabularx}
\usepackage{leftidx}
\usepackage{afterpage}
\usepackage{cprotect}
\usepackage{subcaption}
\usepackage{tikz}
\usepackage{mwe,tikz}
\usepackage[percent]{overpic}

\graphicspath{ {HD50896_plots/} }

\newcommand{\kms}{\hbox{km\,s$^{-1}$}}
\newcommand{\Vinf}{\hbox{$V_\infty$}}
\newcommand{\Rstar}{\hbox{$R_\mathrm{\ast}$}}
\newcommand{\Rsun}{\hbox{$R_\odot$}}
\newcommand{\Msun}{\hbox{$M_\odot$}}
\newcommand{\Lsun}{\hbox{$L_\odot$}}

\newcommand{\Msunyr}{\hbox{$M_\odot \,\hbox{yr}^{-1}$}}
\newcommand{\Mdot}{\hbox{$\dot M$}}

\newcommand{\gl}{$\lambda$}
\newcommand{\glgl}{$\lambda\lambda$}

\newcommand{\Heilineopt}{\ion{He}{I} \gl5877}
\newcommand{\Heiiline}[1]{\ion{He}{II} \gl#1}

\newcommand{\civdoubopt}{\ion{C}{IV} $\lambda\lambda$5801, 5812}

\newcommand{\taueff}{\hbox{$\tau_{\hbox{\scriptsize eff}}$}}

\newcommand{\blankhalf}{\par\vskip 4pt\noindent}

\newcommand{\cmfgen}{{\mbox{\sc cmfgen}}}
\newcommand{\formpol}{{\sc form\_pol}}

\newcommand{\rhosm}{{\ensuremath{\rho_{\hbox{\scriptsize sm}}}}}
\newcommand{\mylist}{\par\indent\hangindent=40pt\hangafter=1}

\newcommand{\tsup}[1]{\textsuperscript{#1}}

\newcommand{\betaav}{\hbox{$\langle \beta \rangle$}}
\newcommand{\betasob}{\hbox{$\beta_{\scriptstyle S}$}}
\newcommand{\betasm}{\hbox{$\beta_{\scriptstyle sm}$}}
\newcommand{\betaed}{\hbox{$\beta_{\scriptstyle ed}$}}

\title[Shell Wind Model of HD~50896]{Using Shell Models to Investigate Clumping in the WN4 Star HD~50896}

\author[Brian L. Flores, D. John Hillier, and Luc Dessart]{
	Brian L. Flores$^{1}$, 
	D. John Hillier$^{1}$\thanks{hillier@pitt.edu}, and Luc Dessart$^2$
	\\
$^{1}$ Department of Physics and Astronomy \& Pittsburgh Particle Physics, Astrophysics and Cosmology Center (PITT PACC), University of Pittsburgh, \\
3941 O'Hara Street,  Pittsburgh, PA 15260, USA.\\
$^2$ Institut d'Astrophysique de Paris, CNRS-Sorbonne Universit\'e, 98 bis boulevard Arago, F-75014 Paris, France.\label{inst2}
}

\date{Accepted XXX. Received YYY; in original form ZZZ}

\pubyear{2015}

\begin{document}
	\label{firstpage}
	\pagerange{\pageref{firstpage}--\pageref{lastpage}}
	\maketitle
	
	\begin{abstract}
	The spectra of Wolf-Rayet (WR) stars exhibit strong, broad emission lines that originate in the wind. These winds are radiatively driven and are susceptible to hydrodynamic instabilities that result in the formation of clumps. When modelling spectra of WR stars the volume-filling factor (VFF) approach is usually employed to treat clumpy winds. However, it is based on the assumption that the entire wind mass resides in optically thin clumps, which is not necessarily justifiable in dense winds. 
	To test the validity of the VFF approach we use a previously described method of treating clumping, the ``Shell'' approach, to study line and continuum formation in the dense wind of the WN4  star, HD~50896. Our models indicate that fully intact spherical shells are in tension with observed spectra; a persistent ``dip'' in emission lines  occurs at line centre. Removing this dip requires our models to use  ``broken'' shells -- shells that are highly decoherent laterally. This insinuates that the wind of HD~50896, and by extension the winds of other WR stars, are comprised of small laterally confined and radially compressed clumps -- clumps smaller than the Sobolev length. We discuss some of the conditions necessary for the VFF approach to be valid. 
	\end{abstract}
	
	
	
	
\section{Introduction}\label{sec_intro}
	Wolf-Rayet (WR) stars, a class of  massive stars descended from stars greater than 20\,\Msun, are characterized by the presence of strong broad emission lines in their spectra. The presence of these emission lines is indicative of a strong outflow; their wind densities are typically an order of magnitude higher than those in massive O stars. And just like their O star progenitors, evidence for highly clumped winds is abundant. Observationally, variability seen in line profiles \citep{Moffat1988,Robert1994,Lepine2000} and polarization \citep{Drissen1987,StLouis1987,StLouis1993,Brown1995} provides evidence of small-scale structures or ``blobs'' propagating outward in the winds. Theoretical support for clumped winds comes from the theory of time-dependent radiation-driven winds \citep{Owocki1984,Owocki1988,Owocki1994,Feldmeier1995}. Taking into account the effects of clumping in radiative transfer codes is pivotal for deriving accurate mass-loss rates of WR stars, and is also potentially important for accurate parameter and abundance determinations.
	
	\cite{Hillier1991} provided one of the earliest insights into the spectroscopic effects of clumping -- specifically, a reduction in strength of the electron scattering wings of emission lines compared to non-clump models. In the models with clumping, a shell-like approach was used, with an effective volume-filling factor (VFF) of approximately 0.5. These results confirmed ideas that similar
	spectra, but with weaker electron scattering wings,  could be obtained by using clumped wind models with a lower mass-loss rate.
	An inherent weakness of the approach is the need for a large number of grid points, which greatly increases the computational effort.
	
	In later work a volume-filling factor (VFF) approach was adopted \citep{Hillier1996,Hamann1998,Hillier1999} since this has the same computational requirements as an un-clumped model. In the VFF approach, the wind is assumed to have a homogeneous distribution of optically thin clumps that occupy a fraction of the volume, $f_{V}$, with a void interclump medium (ICM). With these assumptions, $\rho$-dependent features remain unchanged while $\rho^2$-dependent features are enhanced by the inverse of $f_{V}$. For WR stars, the VFF approach provides a means to simultaneously fit the electron scattering wings and most emission lines and continua \citep{Hamann1998,Hillier1999}. However, it is unclear whether the assumptions (e.g. clumps are optically thin) used to justify the VFF approach are valid. In the spirit of the VFF approach, approximate techniques have also been developed to treat  porosity and vorosity \footnote{Vorosity refers to porosity in velocity \citep{Owocki2008} -- it arises from the introduction of gaps in velocity spaces and affects line transfer.} \citep{2014A&A...568A..59S}, and these have been used to investigate clumping and porosity in the winds of  O stars \citep[e.g.][]{Sundqvist_Puls_2018,Hawcroft2021}.
	
	To accurately model the clumpy stellar winds of massive stars would require a 3D non~LTE radiative transfer code that is capable of handling shocks and non-monotonic flows. However, calculations with such codes would be computationally expensive.  Consequently workarounds are implemented to model clumpy winds of WR stars. One method, implemented in works by the Potsdam group \citep{Feldmeier2003,Oskinova2004,Oskinova2007}, is to assume that the clumps are highly compressed, almost ``pancake-shaped" fragments, when studying X-ray emission lines originating from strong winds.

	In our previous paper \citep[][hereafter Paper I]{Flores2021} we describe a method for treating clumping without assuming optically thin clumps and a void ICM. The method described therein uses highly dense, spherically symmetric shells embedded on a non-void ICM to treat clumpy winds; this is called the ``Shell'' method. While very idealized, it allows for clumping, the interclump medium, and vorosity. We used the ``Shell''  method in Paper I to model the clumpy wind of AzV83, an O7Iaf+ supergiant in the Small Magellanic Cloud, and made comparisons to a VFF model in the 1D radiative transfer code \cmfgen, and found that for this star the Shell and VFF approaches yielded similar results. While our studies provide insights into the effects on ionization structure and spectral characteristics of AzV83 when using the Shell approach, a number of questions still remain: In what regimes does the use of coherent (i.e.\ spherically symmetric, laterally-coherent) shells introduce artifacts into the modelling that affect derived parameters? In what regimes does treating winds with the Shell approach perform worse compared to the VFF approach, and why? 
	
	In this study we continue our investigation of using shells to represent clumps and discuss the differences in spectra compared to those computed using the VFF approach. For this paper, we choose models that are appropriate for the galactic WN4 star, HD~50896 (also known as a WR 6 and EZ CMa). HD~50896 is one of the most well-studied WR stars and is considered a prototypical WNE star \citep{Hillier1988}.

	HD~50896 is spectroscopically and photometrically variable \citep{Wilson1948,Ross1961} on a time scale of 3.76 days \citep{Firmani1980}, and it is a strong X-ray emitter ($L_X  = 1.1 - 11.2 \times 10^{32}$ ergs s$^{-1}$, \citealt{Skinner2002})\footnote{The distance to HD~50896 is uncertain with distance estimates ranging from $d = 0.58$ to $2.3$ kpc. The Gaia EDR3 release gives a parallax of
	$0.650\pm 0.037$ mas, which is significantly larger than the value of $0.441\pm 0.065$ mas given in DR2 \citep{Crowther2020}. L$_X$ is the unabsorbed (i.e. corrected for interstellar absorption via spectral fitting) luminosity in the 0.2–10 keV range.}. It has been suggested that this variability is due to binarity \citep{Firmani1980}. The observed variations could be explained by the companion star's orbital motion \citep{Morel1999,Moreno2005,Toledano2007} or wind-wind collision, if a strong stellar wind is present \citep{Marchenko1997,Flores2001}. Such effects are common in other binary systems \citep{Stevens1999,Chevrotiere2011}. However, the binary model for HD~50896 was generally discounted because  non-coherent phase-dependent variability over a timescale longer than a couple of weeks has been observed \citep{Drissen1989,Robert1992}.
	
	The presence of a compact companion has also been suggested to explain the observed variations, although this idea is not compatible with observations \citep{Stevens1988,Pollock1989,Skinner1998}. Other suggestions have included the presence of co-rotating interaction regions (CIRs), possibly due to magnetic spots \citep{StLouis2018}. The observed period would then be set by the rotation rate, while changes in the CIR footprints would allow for variations in phase.
	
	More recently it has been suggested that the phase changes in HD 50896 are caused by apisidal motion \citep{Skinner2002,Schmutz2019}, thus re-raising the possibility that it is a multiple star system. \citet{Koenigsberger2020} suggested that the system has a eccentricity of 0.1, and that the companion is a late B-type star. Such a system can explain the observed radial velocity variations and the X-ray light curve \citep{Koenigsberger2020}.
     
	For our modelling, we will ignore the companion star since the variability, while persistent, only has a small effect on line profiles \citep[e.g.][]{Flores2007,Flores2011} and photometry \citep{Schmutz2019}, and will not influence any conclusions reached by our analyses.
		
	This paper is organized as follows. In Section \ref{sec_obs_mod} we briefly present spectroscopic observations of HD~50896 and outline the two approaches, VFF and Shell, for treating clumpy winds. In Section \ref{sec_results} we compare and discuss the atmospheric and wind models of the two treatments, emphasizing the challenges of using shells to treat the clumpy wind of HD~50896. A comparison between the VFF and Shell approaches is made in Section~\ref{shell_vs_vff}, while in Section~\ref{sec_esc_mc} we address under what conditions the VFF approach is valid.  In Section \ref{sec_discuss}, we discuss the effectiveness of using shells, and highlight faults and benefits compared to the VFF approach. Conclusions and future work are presented in Section \ref{sec_conclusion}.
	

\section{Observations and Modelling}\label{sec_obs_mod}


	\subsection{Observational Data}
	\label{subsec_obs}
	
	
	For our study we utilized both UV and optical data. The optical \ion{He}{ii} $\lambda$5411 and $\lambda$4687 profiles were obtained by Hillier in the mid 1980s using the Anglo-Australian Telescope. The other optical data set is from \cite{Torres-Dodgen1988,Torres-Dodgen1999}.	The UV spectrum is from \cite{Howarth1986} and was kindly supplied to one of the authors (Hillier) by Ian Howarth. It is a high-signal-to-noise spectrum that was created from 58 IUE	spectra. For the optical data we used the low-resolution photometric spectrum of \cite{Torres-Dodgen1986}.
	
	
\subsection{Atmospheric and Wind Modelling}
	\label{subsec_mod}
	
	
	The method used to study HD~50896 is similar to that used in Paper I 
	for AzV83. \cite{Flores2021} used the ``Shell'' option  in the non~LTE line-blanketed multi-purpose atmospheric code \cmfgen\footnote{The Shell approach will become publicly available in a future update for \cmfgen\ on \url{http://www.pitt.edu/~hillier}. Register on the website to receive emails about updates to \cmfgen.}  \citep{Hillier1998} to undertake spectroscopic studies of AzV83. In this paper we adopt the same basic technique for the non~LTE calculations. Consequently only the basic assumptions are briefly described here -- the reader should access the original paper for further details. Some changes to the	spectral computations have been made, and these are described in the appropriate sections.
	
	We constructed two sets of models to investigate the difference
	in spectra obtained with the Shell and VFF approaches. The models used similar atmospheric parameters ($L_* = 3.0\times 10^5 \Lsun$, $R_{*} = 2.5 \Rsun$, $T_{\rm eff} = 48,430 $K, \Vinf=2000\,\kms, and $\Mdot = 3.5\times10^{-5} \Msunyr$ ), atomic models, and surface chemical abundances (listed in Appendix \ref{appendix:A1}). We do not solve the hydrodynamic equations for the wind; instead we assume that the wind's velocity field is described by the $\beta$ velocity law. This velocity law differs from that previously used by \cite{Hillier1988}, but it does not change the results significantly. Based on earlier work, we adopt a double beta law with $\Vinf = 2000 $ \kms. The double velocity law has $\beta=1$ in the inner wind, but a slower approach to $\Vinf$ in the outer wind \citep{Hillier1999}. To limit the computational effort we included only 6 atomic species, and used many fewer ionization stages than we would normally use when modelling WR spectra. While this may affect agreement between model spectra and observations, it will not affect any conclusions which are only concerned with differences between spectra computed with the two different clumping approaches.

	The assumption of a smooth velocity field in this work is a necessary approximation, but is at odds with simulations which show very complex velocity fields \citep[e.g.][]{Owocki1988}. However, while our velocity law is continuous in the Shell model, we can still get vorosity effects because of the lower densities in the interclump medium. The VFF approach also ignores the shocked wind, but X-rays arising from these shocks are included (via several free-parameters that describe the shock temperature, the total X-ray luminosity, and the emission distribution in the wind).
	
	In \cmfgen, the VFF approach uses an ad-hoc, simple parametric treatment for the volume-filling factor $(f(r))$: 
	\begin{equation}
		f(r)= f_{V} + (1-f_{V})\cdot \exp[-\varv(r)/\varv_{\rm cl}]\,\,,
		\label{eq:fcl}
	\end{equation}
	where $f_{V}$ is the volume-filling factor at infinity and $\varv_{\rm cl}$ is the characteristic velocity scale. In this method, the wind is assumed to be homogeneous close to the star but becomes clumpy, with a volume-filling factor $f_{V}$, at large distances. The volume-filling factor in the VFF approach was set at 0.1 and $\varv_{\rm cl}$ to 100\,\kms . Alternative formulae can be adopted, but in general it is observationally difficult to constrain additional free parameters in both O and WR stars \citep[e.g.][]{Najarro2011,Hillier1999,Aadland2021}. There is some evidence that $f_V$ increases at larger radii in O stars \citep[e.g.][]{Puls2006,Rubio-Diez2021}.

\begin{figure} 
		\includegraphics[height=\columnwidth,angle=-90,trim={2.5cm 0cm 0cm 0cm},clip]{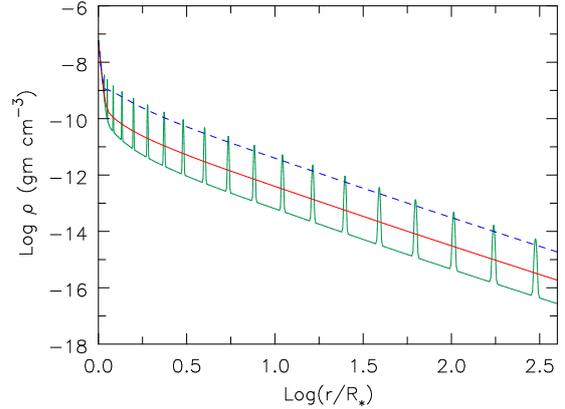}
		\caption{Comparison of the wind density distributions used in modelling HD~50896. The red solid curve is the smooth wind density distribution, the blue dashed curve corresponds to the VFF model with $f_{V}=0.1$, and the green solid curve is the Shell model. }
		\label{fig_1_dens}
	\end{figure}

	For the Shell approach we explicitly insert Gaussian-like ``shells'', logarithmically spaced  (approximately) in radius, into the wind. To define the shell grid we first compute a characteristic shell template -- a  low density background on which a dense shell (typically Gaussian in shape) is added. The parameters used to create the template are the length of the template ($L_t$), the Full-Width Half-Maximum (FWHM) of the dense shell ($W_{\rm FWHM}$), the minimum background level ($M_{\rm BL}$), and the number of depth points ($ND\textsubscript{t}$).  The peak density of the Gaussian shell is then adjusted to yield the same mass for the structure as a structure of constant unit density. This basic template is then used to define the shell structure of our model. The template can be stretched (we usually have an increased shell density at low velocities), and we also modify the clumping at low velocities so the effective volume-filling factor goes to one. As for the VFF approach, we choose an onset velocity ($V\textsubscript{onset}$; typically  30\,\kms), and a characteristic density contrast scale (V\textsubscript{clump}; typically 100\,\kms). Figure \ref{fig_1_dens} plots the wind density distribution for the Shell (with $L_t=2$, $W_{\rm FWHM}=0.05$, and $M_{\rm BL}=0.01$) and VFF models. Both models have a volume-filling factor of 0.1.

 	For all our Shell models the ratio of the density in the clumps to that in the inner clump region is large (i.e. $\approx 100$). As shown in Appendix~\ref{app_f_constaint}, a large value is required if our Shell model is to have an effective VFF of 0.1 (typically found/used in WR modelling). With a VFF of 0.1 we reduce the strength of the electron scattering wings in WR stars, compared to a smooth wind model, by approximately a factor of three. For $f=0.1$, the ratio must  be in excess of 40. Provided this constraint is met, we can choose an arbitrary density contrast.
	 
\section{Results}
	\label{sec_results}
	
	
\subsection{Ionization Structure}
	\label{subsec_ion_structure} 
	
	
	It has been shown previously that the luminosity has an indirect influence on the strength of \ion{He}{I} and \ion{He}{II} lines.	This is due to the luminosity's influence on the ionization structure of helium. As discussed by \cite{Hillier1987}, the ionization of He$^+$ primarily occurs from the $n=2$ level,	and is controlled by the photoionizing photons ($\lambda < 912$\,\AA) and the radiation field in the \ion{He}{ii} Ly$\alpha$ transition. An increase in luminosity moves the radial location at which He$^{+}$ becomes the dominant ionization stage further out, decreasing the strength of neutral helium recombination lines.

\begin{figure}  
		\centering
		\includegraphics[scale=0.6,trim={2.1cm 2cm 6cm 16.5cm},clip]{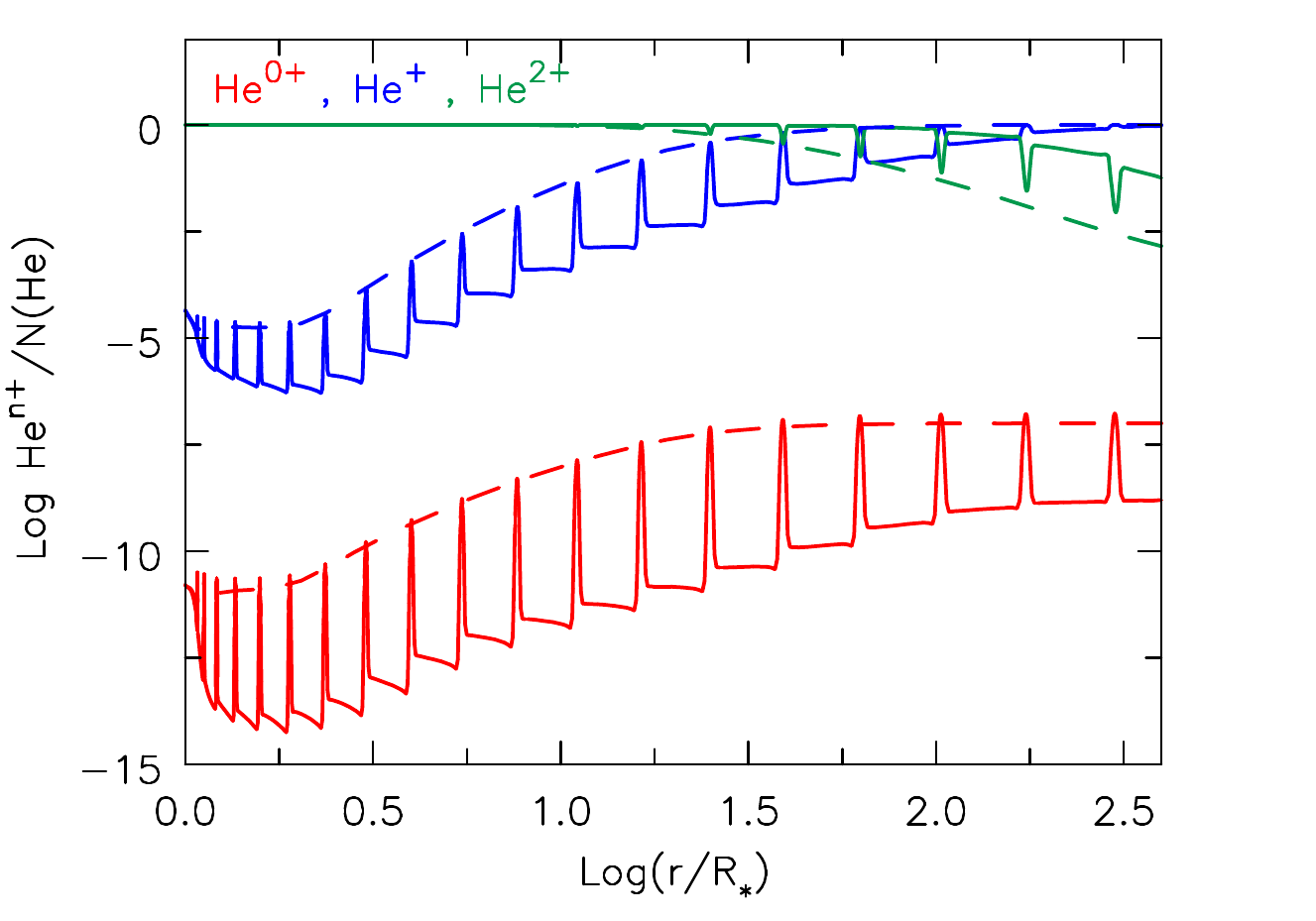}
		\includegraphics[scale=0.6,trim={2.4cm 2cm 6cm 16.5cm},clip]{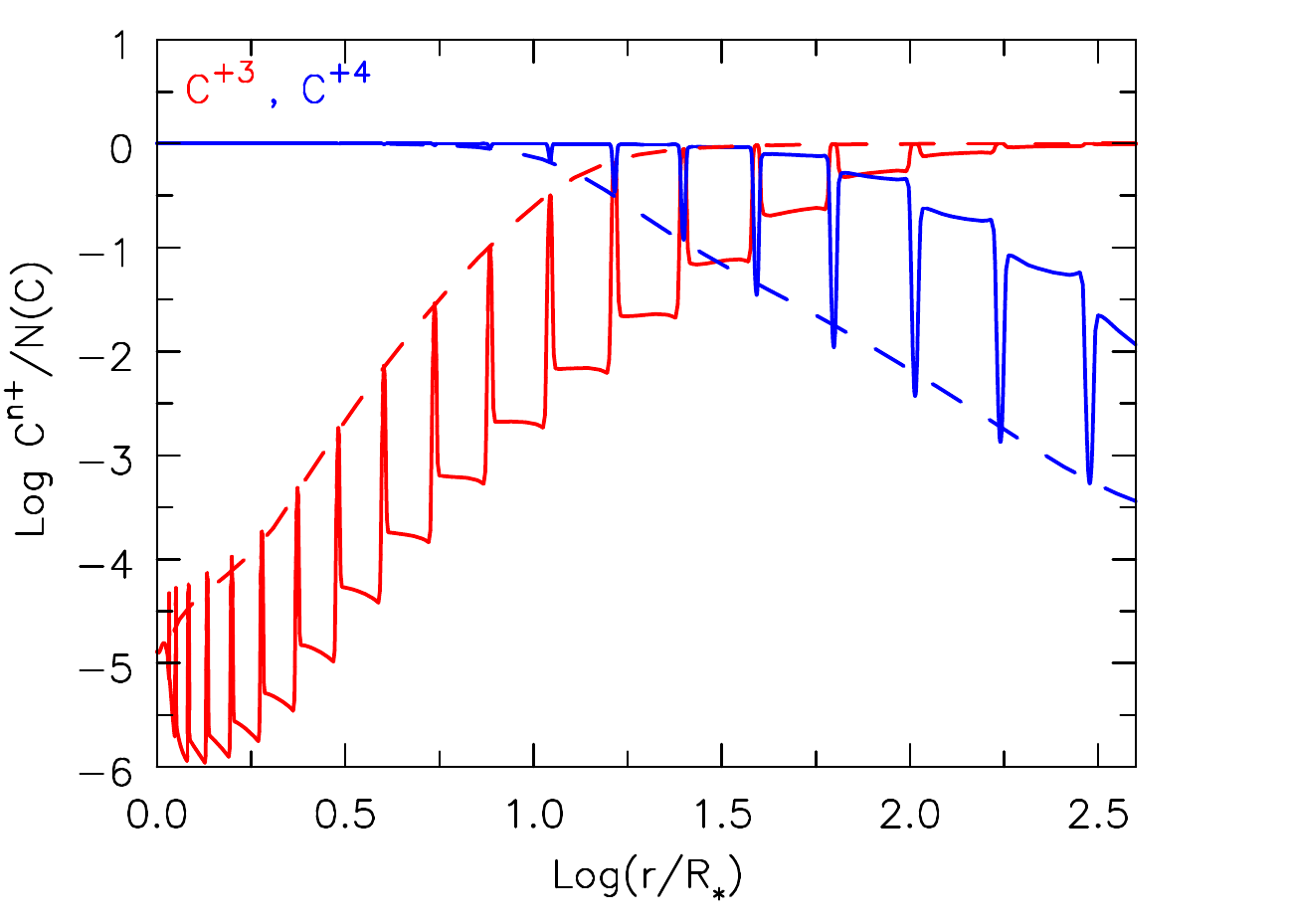}
		\caption{Plots of the ionization structure of helium and carbon in both the Shell model (\textit{solid line}) and VFF model (\textit{dashed line}). The ionization change seen in the Shell model, specifically within the shells, mimics that seen in the VFF model -- at large radii, the dominant ionization state of helium and carbon shifts from He$^{++}$ to He$^{+}$ and from C$^{4+}$ to C$^{3+}$, respectively. 
		}
		\label{fig_He_IF}
	\end{figure}

	Figure \ref{fig_He_IF} shows the ionization structure of helium and carbon in the wind of HD~50896 for a Shell and a VFF model. The helium ionization structure in the Shell model -- specifically within the shells -- qualitatively follows the same radial trend seen in the VFF model. Importantly, the dominant ionization state of helium shifts from He$^{++}$ to He$^{+}$ at approximately the same radius. As a consequence, we should expect the ratio of \ion{He}{II} to \ion{He}{I} line fluxes in the Shell model to be similar to that in the VFF model. The difference in ionization between the shell and interclump medium is large -- one to two orders of magnitude. Such variations, for example, can have a significant impact on the formations of lines from high-ionization species such as \ion{O}{VI} in O stars \citep{Zsargo2008,Flores2021}. The two-phase ionization and level population structure seen in the Shell model cannot be treated using the extended clumping formulation of \cite{Sundqvist_Puls_2018}.

\subsection{Departure Coefficient}
	\label{subsec_DC} 
	
\begin{figure*} 
		\centering
		\includegraphics[scale=0.33,angle=-90,trim={2.5cm 1cm 0cm 2cm},clip]{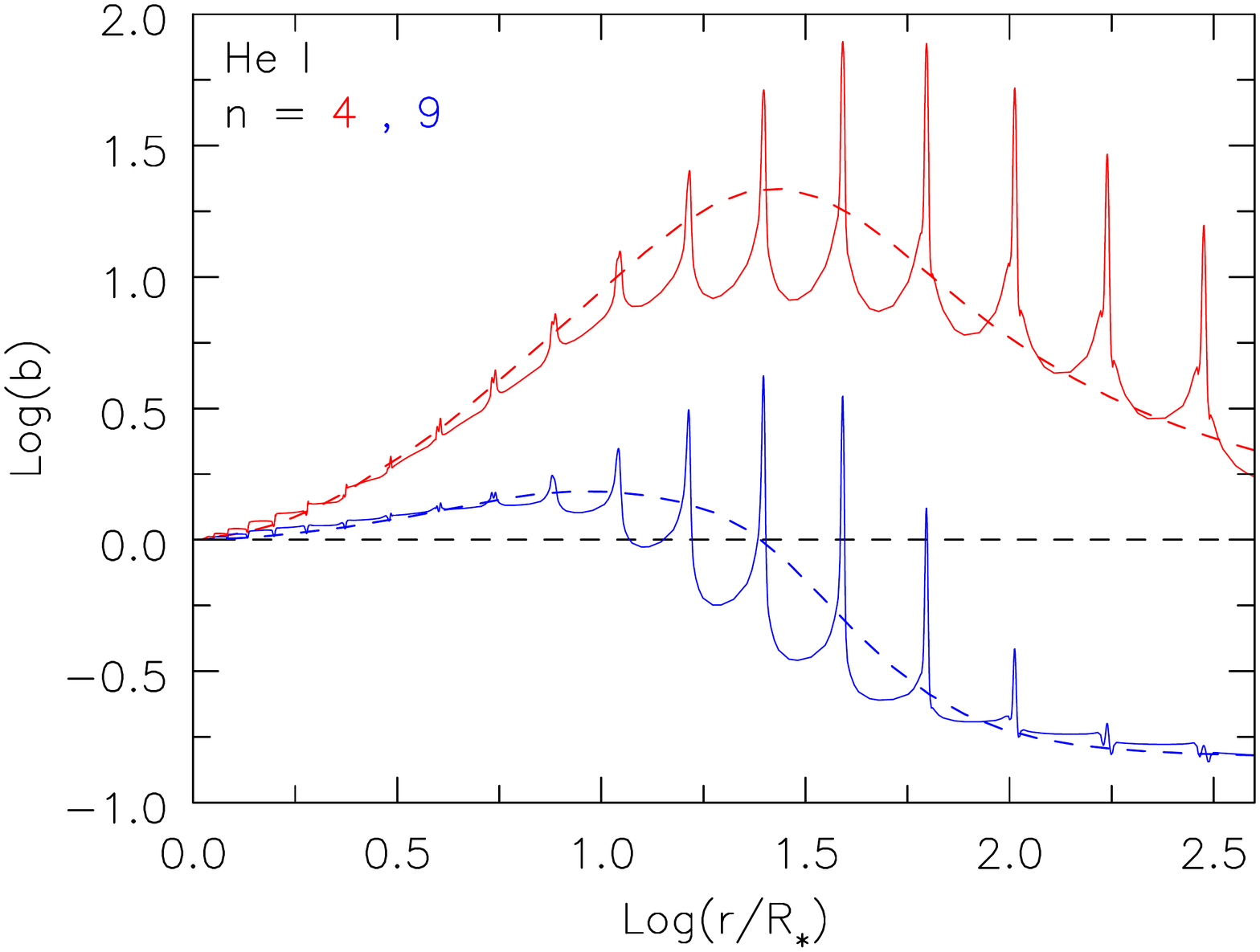}
		\includegraphics[scale=0.33,angle=-90,trim={2.5cm 1cm 0cm 2cm},clip]{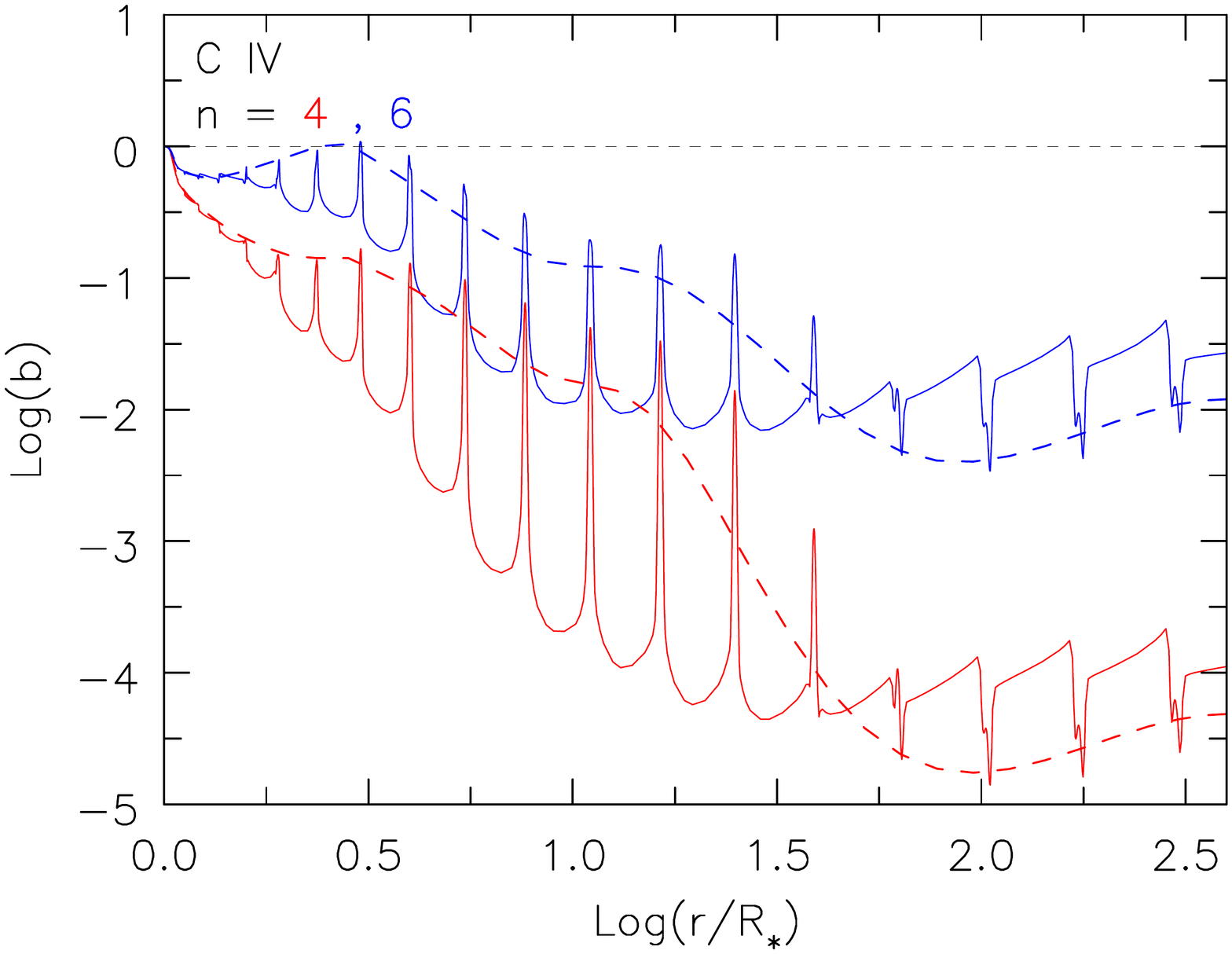}
		\includegraphics[scale=0.33,angle=-90,trim={2.5cm 1cm 0cm 2cm},clip]{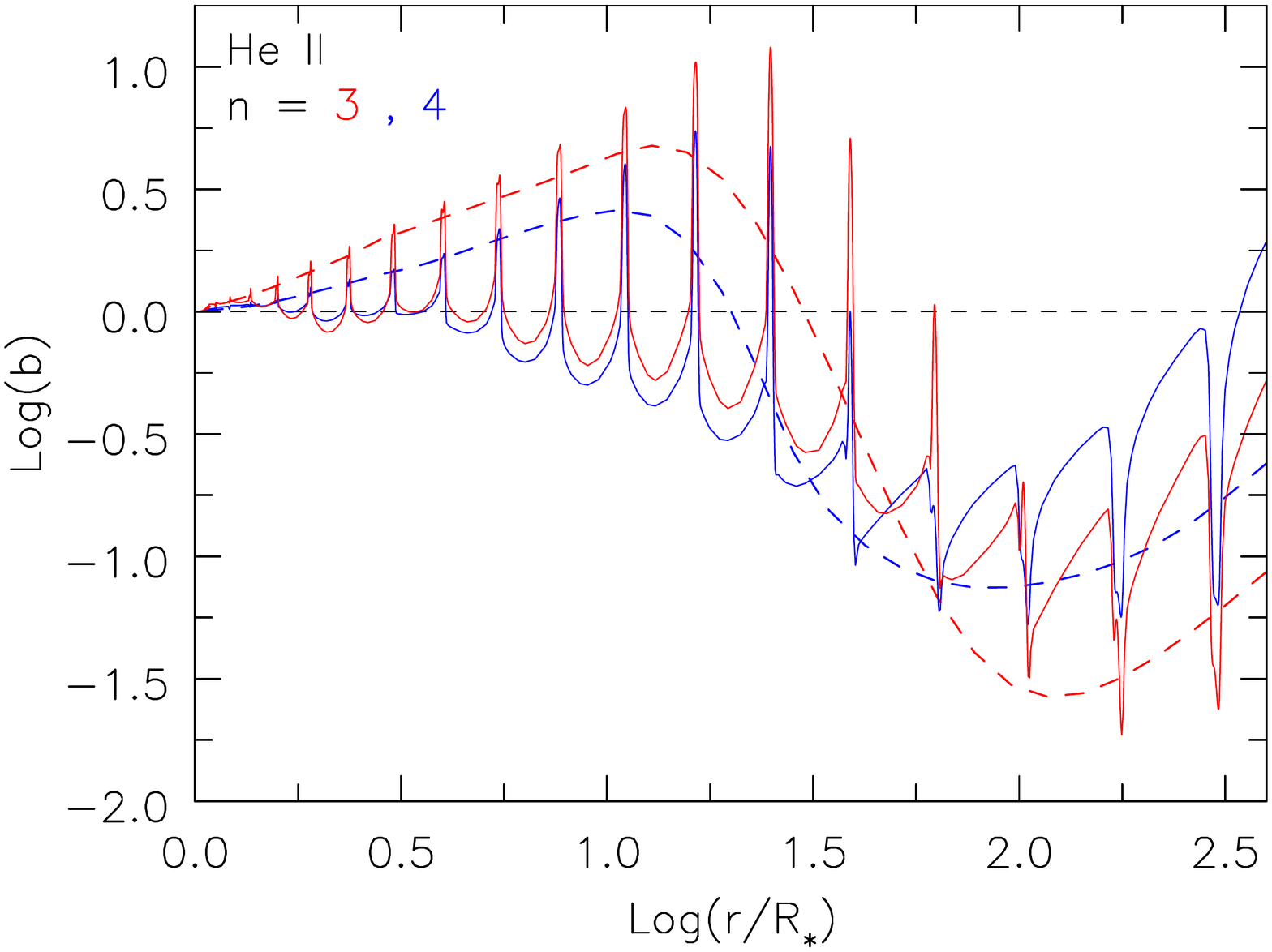}
		\includegraphics[scale=0.33,angle=-90,trim={2.5cm 1cm 0cm 2cm},clip]{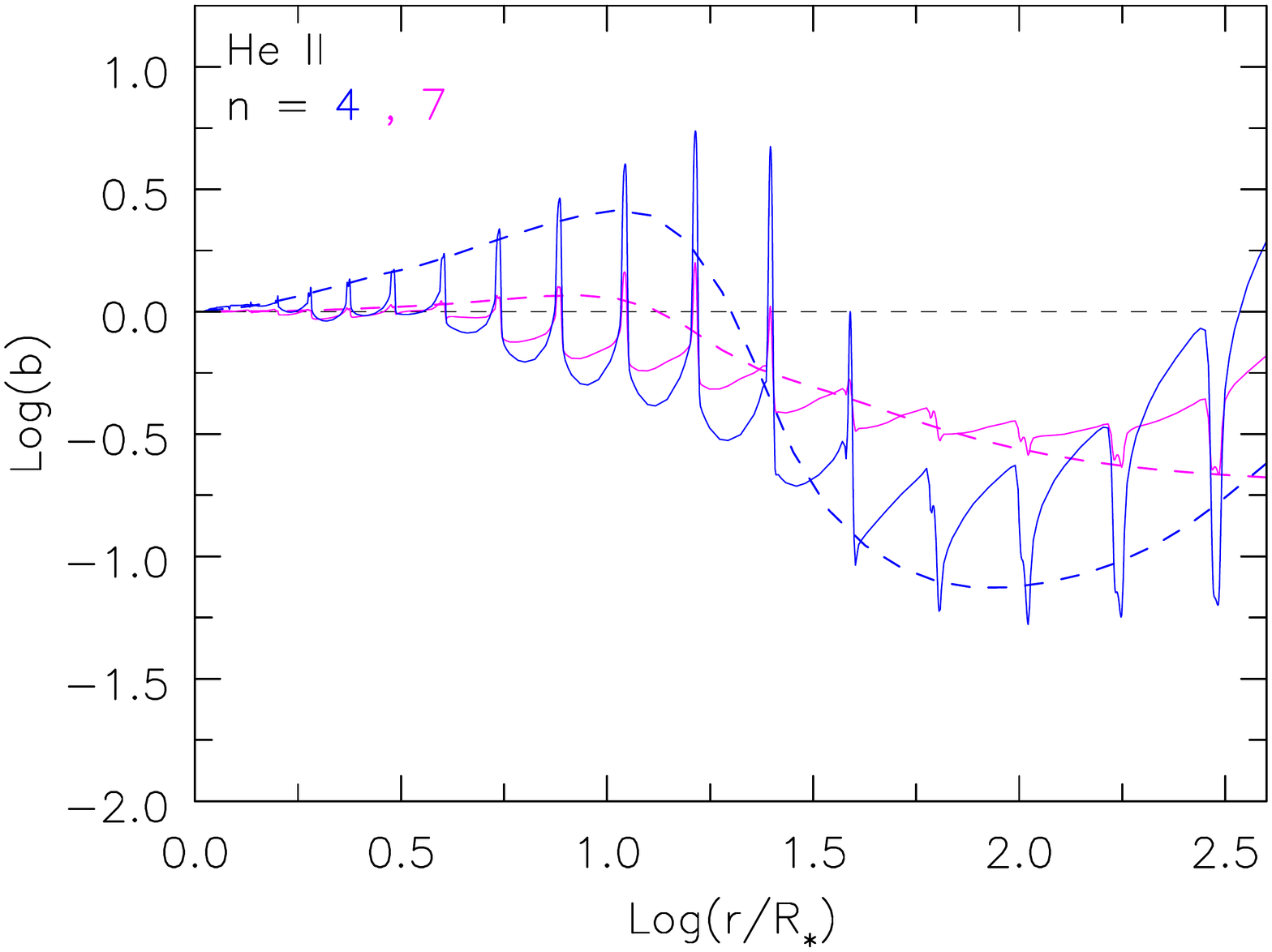}
		
		\caption{Plots of the departure coefficients of selected levels of \ion{C}{IV} (n = 4 and 6; \textit{top right}), \ion{He}{I} (n = 4 and 9; \textit{top left}), and  \ion{He}{II} (n = 3, 4 and 7; \textit{bottom row}) in both the Shell model (\textit{solid line}) and VFF model (\textit{dashed line}). A quantitative discussion of the figures is provided in the text. 
		}
		\label{fig_DC}
	\end{figure*}

	
	To quantify the difference in non Local-Thermodynamic-Equilibrium (non~LTE) results  between the VFF and the  Shell approaches, we plot the departure coefficient ($b$) for several levels of \ion{C}{IV}, \ion{He}{I} and \ion{He}{II} in Fig.~\ref{fig_DC}. These levels are relevant to the formation of several prominent optical emission lines seen in HD~50896: \ion{C}{IV}\ $\lambda5801$ (3d\,\,\tsup{3}D - 2p\,\,\tsup{3}P), \Heilineopt\ (3d\,\,\tsup{3}D - 2p\,\,\tsup{3}P), and \ion{He}{II}\ $\lambda\lambda 4687, 5413$.  Hereafter, we will refer to \ion{C}{IV} electronic states 2p\,\,\tsup{2}P\tsup{o}\  and 3d\,\,\tsup{2}D as levels 4 and 6 respectively. Similarly, we will refer to the \ion{He}{i}\ electronic states 2p\,\,\tsup{3}P\tsup{o}\ and  3d\,\,\tsup{3}D as levels 4 and 9 respectively. 
	
	At first glance, it is evident that the behaviour of the departure coefficients (DCs) is heavily influenced by the clumping treatment implemented in the model; the VFF model shows a smooth variation in DCs, while the Shell model has a ragged  structure. The latter is simply a direct consequence of the large variations in density between the clumps and the interclump regions. It is also evident that the same radial trends are seen in both classes of models, and that in the inner region LTE is achieved. The DCs in the interclump medium (and the temperature), largely irrelevant for the lines discussed in the present paper, are likely to be different in a realistic wind from those shown here -- the interclump (intershell) medium is likely to be strongly influenced by dynamics and associated shocks.
	
	At a given radius the population of a given level can be influenced by many processes such as collisional excitation and deexcitation, photoionization and recombination, photon	escape in lines, and pumping of levels via other lines or the continuum. Consequently, and especially in the inner wind where conditions are close to LTE, the crucial processes are not always easily determined. Thus the following discussion about the behaviour of the departure coefficient should be regarded as illustrative only and a discussion of the many nuances is not included. Further, for some species and/or levels the use of DCs is somewhat meaningless because the levels are mainly coupled to the radiation field, and have only a weak coupling to the local electron temperature.
	
\subsubsection{\ion{He}{II} (n = 3, 4, and 7)} 
	\label{subsubsec_HeII}
	
	 The DCs for several \ion{He}{II} levels are shown in Fig. \ref{fig_DC}. In the inner region, where the radiation is Planckian and collisions are important, the DCs are driven to unity. As we move outward, line transitions remain optically thick while the radiation field becomes diluted. As a consequence (and for the levels shown) the DCs rise above unity. As the density drops, photon escape in lines becomes important, causing the DCs to decline. Eventually the DCs should approach the values determined by simple recombination theory, although pumping via the continuum can delay or inhibit this \citep{Hillier1987}.
	
	The DCs for the Shell model inside $\log(r/R_*) > 1.5$ are greater in the dense clumps, and lower in the interclump regions. This is attributable to the influence of lines. In the innershell region line photon escape is enhanced, while in the dense shells line photon escape is inhibited. In the outer regions, where He is predominantly singly ionized, the situation is reversed -- the dense shells have smaller departure coefficients. We attribute this to the enhanced importance of continuum pumping by the \ion{He}{ii} Lyman lines in the interclump medium.

\subsubsection{\ion{He}{I} (n = 4 and 9)} 
	\label{subsubsec_HeI}
	
	The behaviour of the \ion{He}{I} DCs is similar to those of \ion{He}{II} (top-left plot of Fig. \ref{fig_DC}). In the inner	region the dilution of the radiation field leads to DCs that exceed unity. As the density drops line transitions become more transparent, and the DCs drop. The drop is delayed for the Shell model because of the higher line optical depths in that model (see Sect~\ref{shell_vs_vff}). However, level 4 (2p\,\,\tsup{3}P\tsup{o}) remains overpopulated as we move towards the outer wind. The 2p\,\,\tsup{3}P\tsup{o}\ state can decay to the ground state via an intercombination transition, or to 2s\,\,$^3$S but these routes are relatively ineffective. In the outer wind, the DCs in the Shell model tend to be larger than in the VFF model, especially for level 4 (2p\,\,\tsup{3}P\tsup{o}), which can be attributed to the larger line optical depths in the Shell model. Conversely, in the outer wind, the DCs in the interclump medium more closely agree with the VFF model, indicating the much weaker influence of line opacity. This is further emphasized in the very outer wind, where we see that the DCs for level 9 are	very similar in both the clumps and interclump medium.

	\subsubsection{\ion{C}{IV} (n = 4 and 6)} 
	\label{subsubsec_CIV}
	
	An extensive discussion on the behaviour of \ion{C}{IV} DCs is given by \cite{Hillier1988}; here we simply provide a basic description. In actuality, for \ion{C}{IV}, the use of DCs when discussing the population of levels 4 (3s\,\,$^2$S$_{1/2}$) and 6(3s\,\,$^2$P$\tsup{o}_{3/2}$) is somewhat meaningless, since these levels have little direct coupling to the ion density and electron temperature (as emphasized by the
	large variation in the departure coefficients).
	
	The population density of levels 4 and 6 for \ion{C}{IV} (top-right plot of Fig. \ref{fig_DC}) are determined by the carbon ionization structure and continuum fluorescence via a transition from the ground state  to level 6 (at 312\,\AA). When this transition intercepts continuum radiation at 312\,\AA, it is statistically  favourable to radiatively decay back down to the ground state compared to the 6--4 ($3p$--$3s$) route -- the branching ratio $A_{6 - 1}/A_{6 - 4}$ is \textasciitilde 150. However, in a strong wind (as in HD~50896) the transition to the ground state is optically thick, allowing the 6--4 transition to be the favourable decay route. 
	
	As seen in Fig.~\ref{fig_DC}, both the VFF and Shell models predict levels 4 and 6 to be strongly underpopulated  -- this is primarily due to the coupling of these levels to the	2s and 2p levels, and because the radiation field is effectively much hotter than the local electron temperature. Overall, there is a striking similarity between the DCs in	the dense shells at the same radial location, and those in the VFF model. In the inner region (where C$^{4+}$ is the dominant ionization state) the DCs are larger in the dense shells, but the opposite is true in the outer region (where C$^{3+}$ is the dominant ionization state).  
	
\subsection{Variation of the DCs within a dense shell}
	
	In Sect.~\ref{subsec_DC} we discussed the radial variation of the DCs. In reality the clumps also have structure -- the escape probability and radiation field will vary within the clump.

\begin{figure*} 
		\begin{minipage}[t]{\linewidth}
		\centering
		\includegraphics[scale=0.33,angle=-90,trim={3.5cm 1cm 1cm 1.0cm},clip]{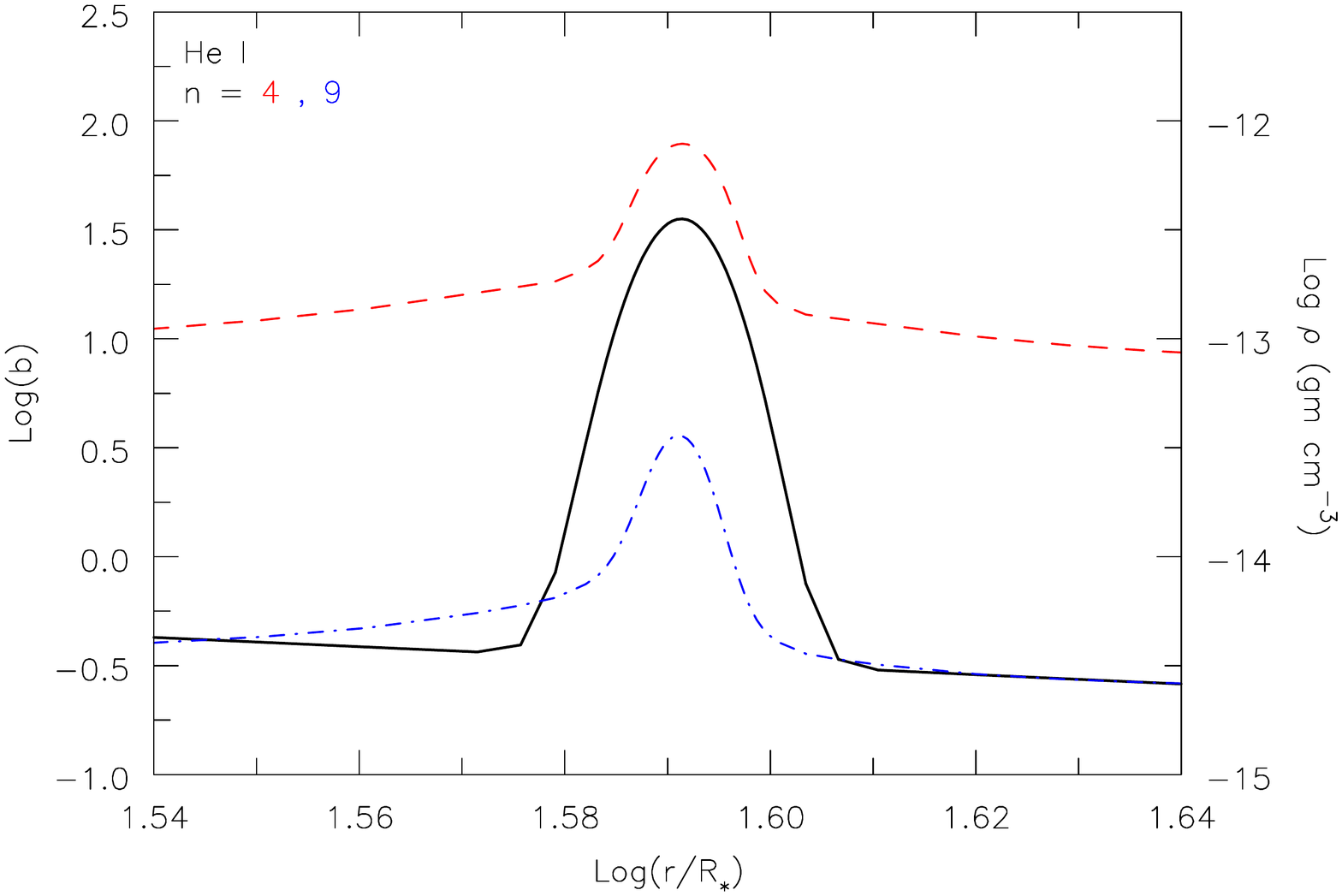}
		\includegraphics[scale=0.33,angle=-90,trim={3.5cm 1cm 1cm 1.0cm},clip]{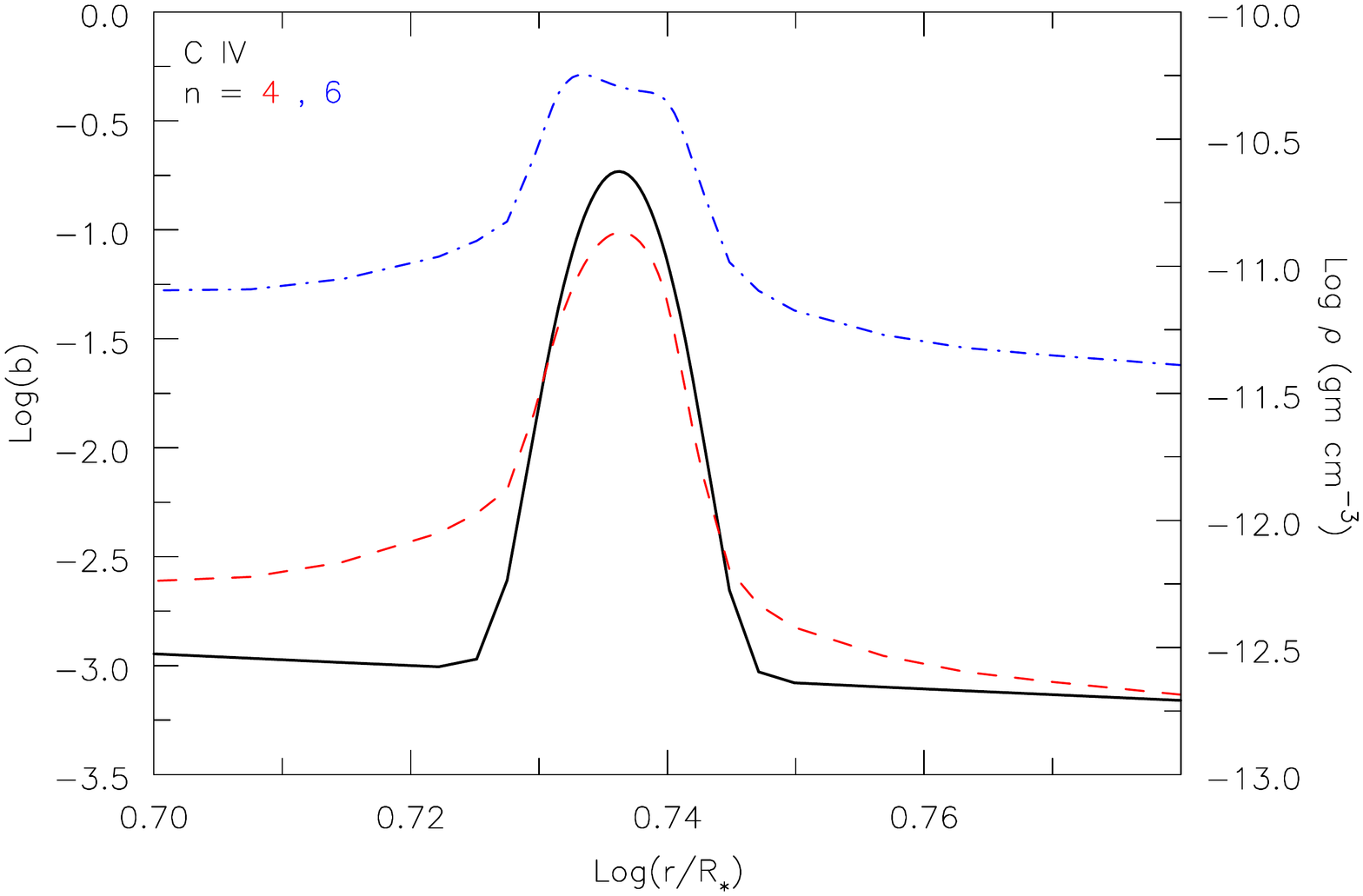}
		\includegraphics[scale=0.33,angle=-90,trim={3.5cm 1cm 1cm 1.0cm},clip]{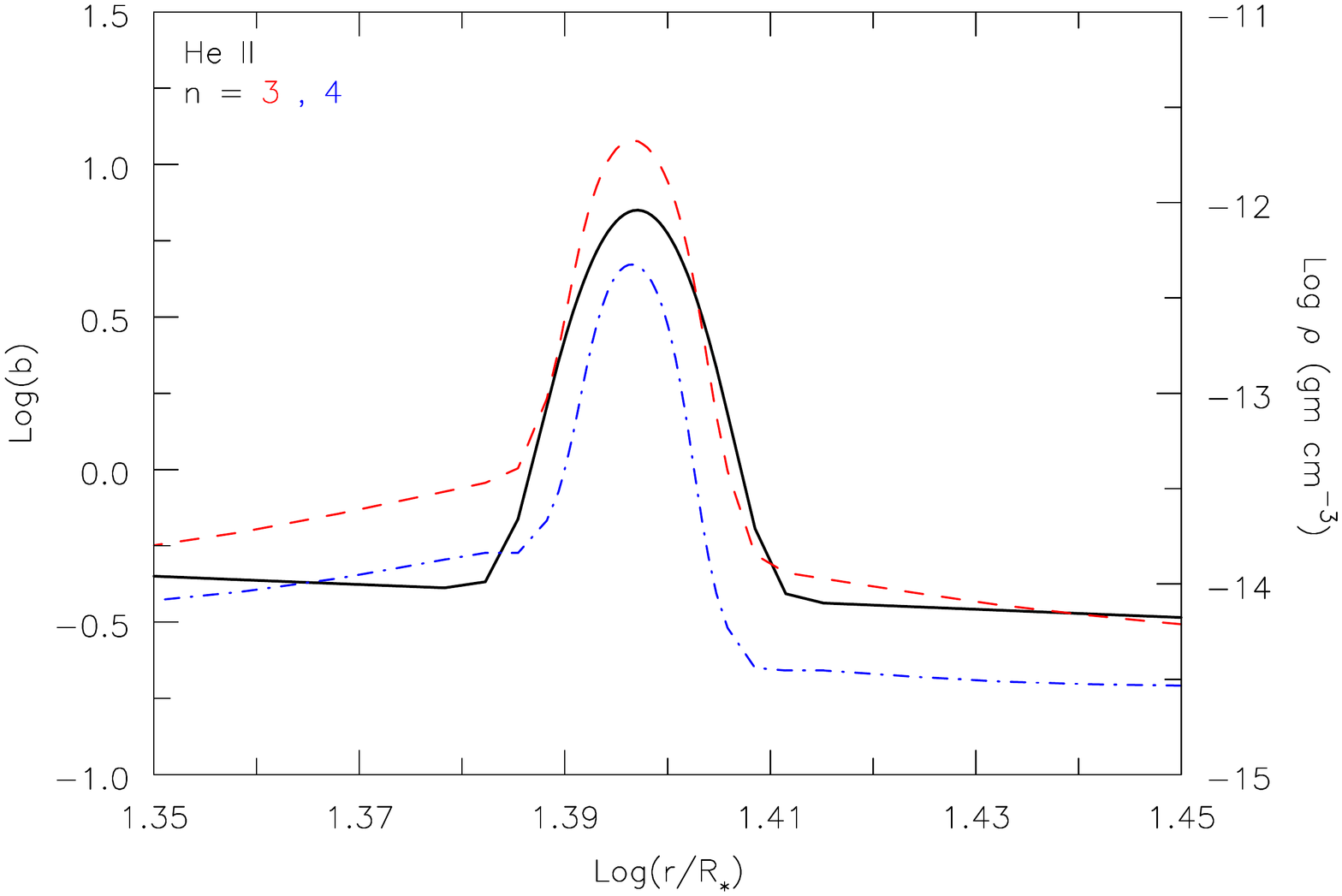}
		\includegraphics[scale=0.33,angle=-90,trim={3.5cm 1cm 1cm 1.0cm},clip]{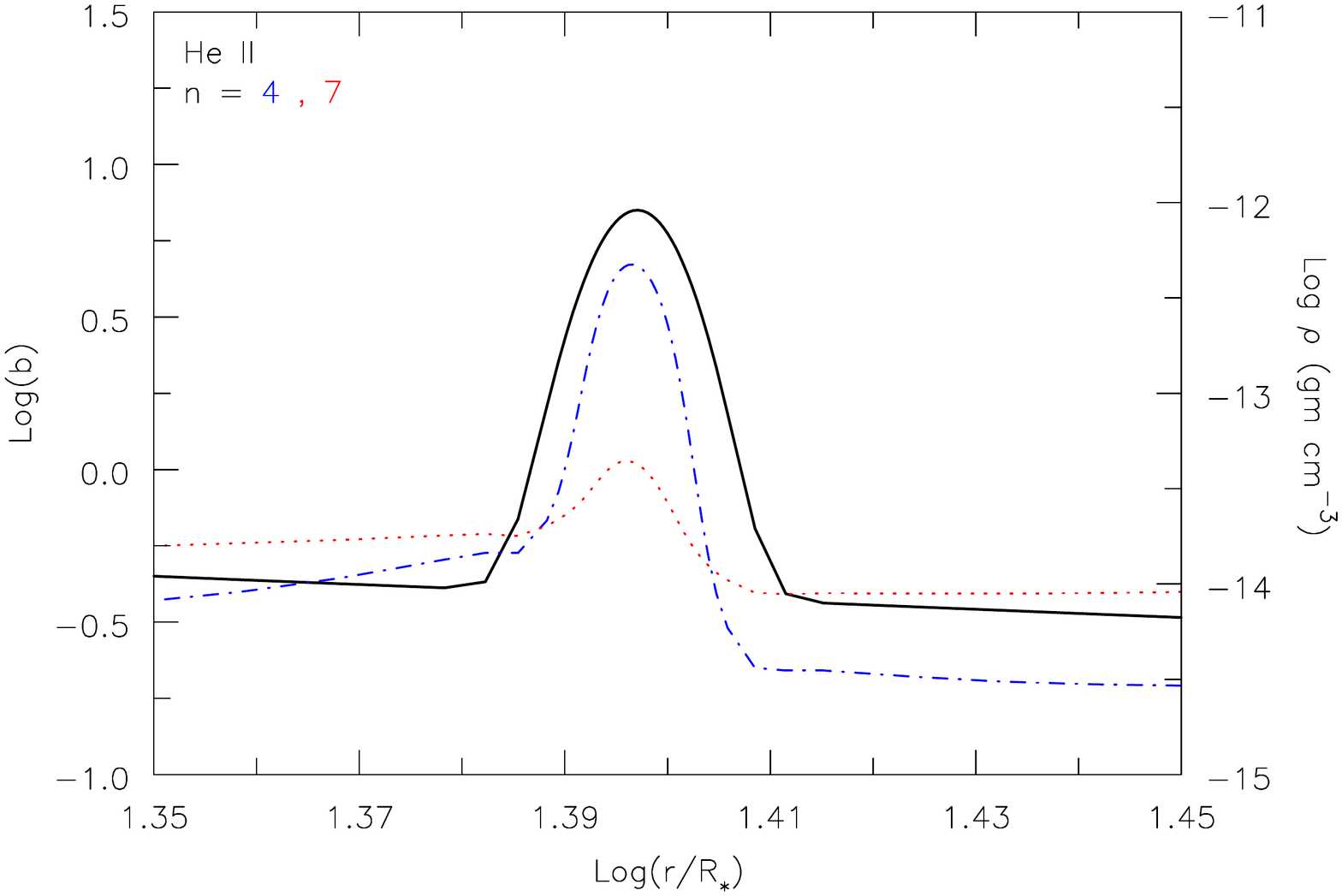}
		\end{minipage}
		\caption{Close-up illustration of the variation in departure coefficients within a clump. Also shown is the density profile (\textit{black, solid curve}); the right-hand axis is the scale for the density profile. 
		}
		\label{fig_DC_close_up}
	\end{figure*}

    In Figure \ref{fig_DC_close_up}, we plot close-ups of the DCs seen in Fig. \ref{fig_DC}, along with the density profile of the wind. We can see that the DC of each species varies across a given shell. The DCs of the upper levels of helium are greater in the core of the shell and fall off towards the shell's edge. Meanwhile, the DC profile of the upper level of \ion{C}{IV} is relatively flat-topped over the entire dense shell. 
    
    Both Fig.~\ref{fig_DC} and Fig.~\ref{fig_DC_close_up} illustrate the complex behaviour that can be exhibited by the DCs in inhomogeneous winds, and help to highlight the difficulty of accurately modelling line profiles formed in winds with complex density structures (and velocity fields).
    
\subsection{Spectral comparisons} \label{subsec_morph}
	
\begin{figure*} 
		\begin{minipage}[t]{\linewidth}
			\centering
			\includegraphics[scale=0.80,trim={0.78cm 2.25cm 0.8cm 1.8cm},clip]{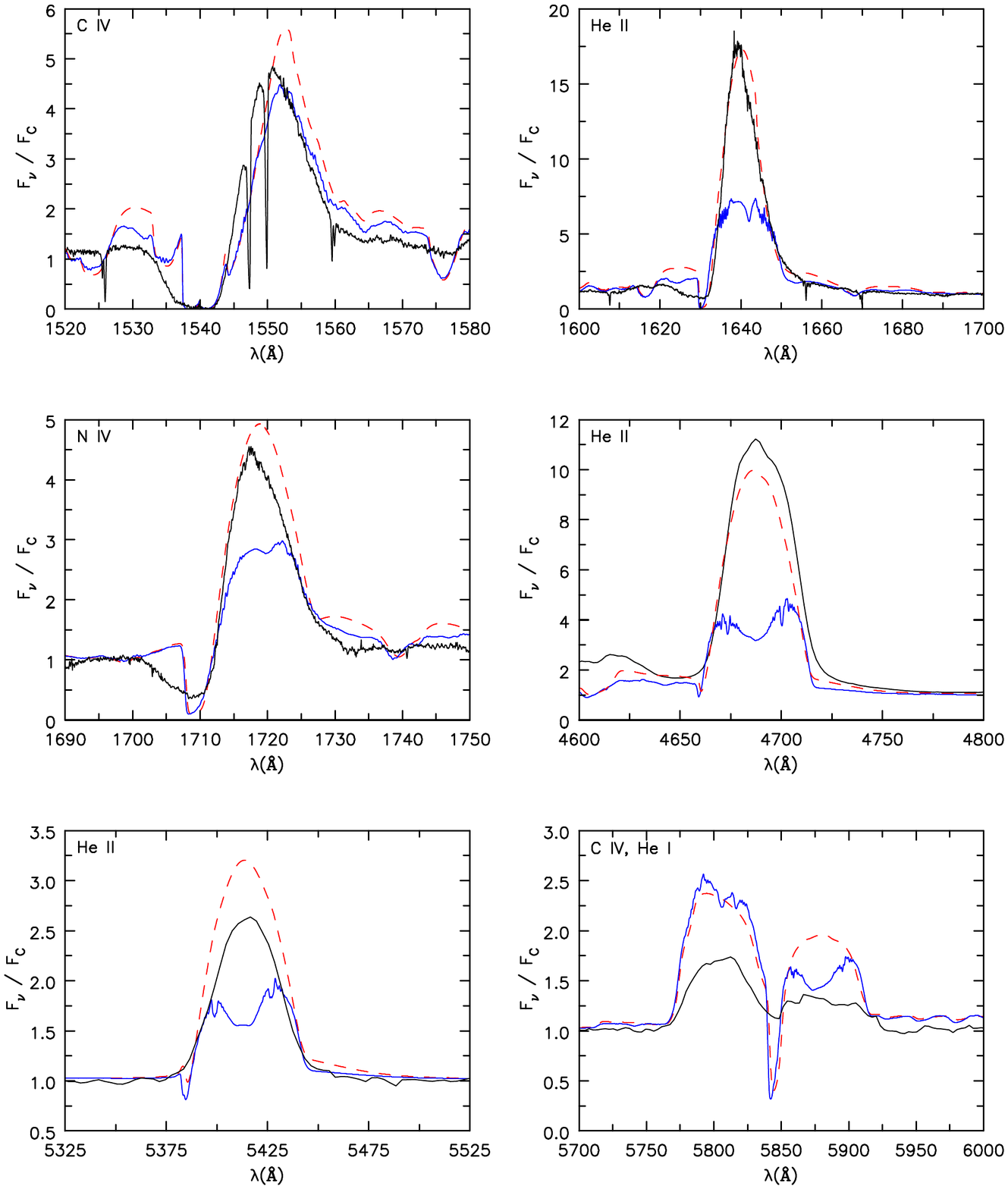}
		\end{minipage}
		\caption{UV and optical spectra of HD~50896 \textit{(black, solid curve)} compared with \cmfgen\ models using VFF approach \textit{(red dashed curve)} and Shell approach \textit{(blue solid curve)}. The Shell models were able to reproduce the UV and optical \ion{C}{IV} lines of the VFF model, but under-predicted other lines, in particular \ion{He}{i}\ and \ion{He}{ii}\ lines where a dip occurs at line centre. This dip is due to an optical depth effect arising from using shells to represent clumps in the wind. 
		}
		\label{fig_spec_comp}
	\end{figure*}

	
    Comparisons of several observed lines for HD~50896 with our model spectra computed using the VFF and Shell approaches (with $\sim 16$ shells) are shown in Fig.~\ref{fig_spec_comp}. From that figure, the VFF model is seen to reproduce several of the lines observed in HD~50896 although some lines are somewhat stronger than observed. The largest discrepancies are for \civdoubopt\ and \Heilineopt. This is partially attributable to the simplicity (i.e. the limited atomic species) in our modelling, and partially attributable to the choice of parameters. However, the most obvious feature illustrated by  Fig.~\ref{fig_spec_comp} is the striking difference between the Shell and VFF line profiles for \ion{He}{I} and \ion{He}{II}.  By contrast the two \ion{C}{IV}\ doublets, \glgl 1548, 1550 and $\lambda\lambda 5801, 5812$, show relatively minor differences.

    The \ion{He}{I}\ and \ion{He}{II}\ lines in the Shell model are weaker than those computed with the VFF model, with \Heilineopt, \Heiiline{4687}, and \Heiiline{5413} showing a central depression. In subsequent sections we explore the cause of the weaker lines and the central depression. To help focus our investigation, we restricted our studies to the optical spectral region of HD~50896, specifically the wavelength range of 5000 -- \SI{6000}{\angstrom}, which features \Heiiline{5413}, \civdoubopt, and \Heilineopt. This set of lines is influenced by different formation mechanisms [i.e. recombination (\Heiiline{5413} and \Heilineopt) and continuum fluorescence (\civdoubopt)] and/or optical depth effects.

\begin{figure*}  
		\centering
		\includegraphics[scale=0.33,angle=-90,trim={2.5cm 1cm 0cm 2cm},clip]{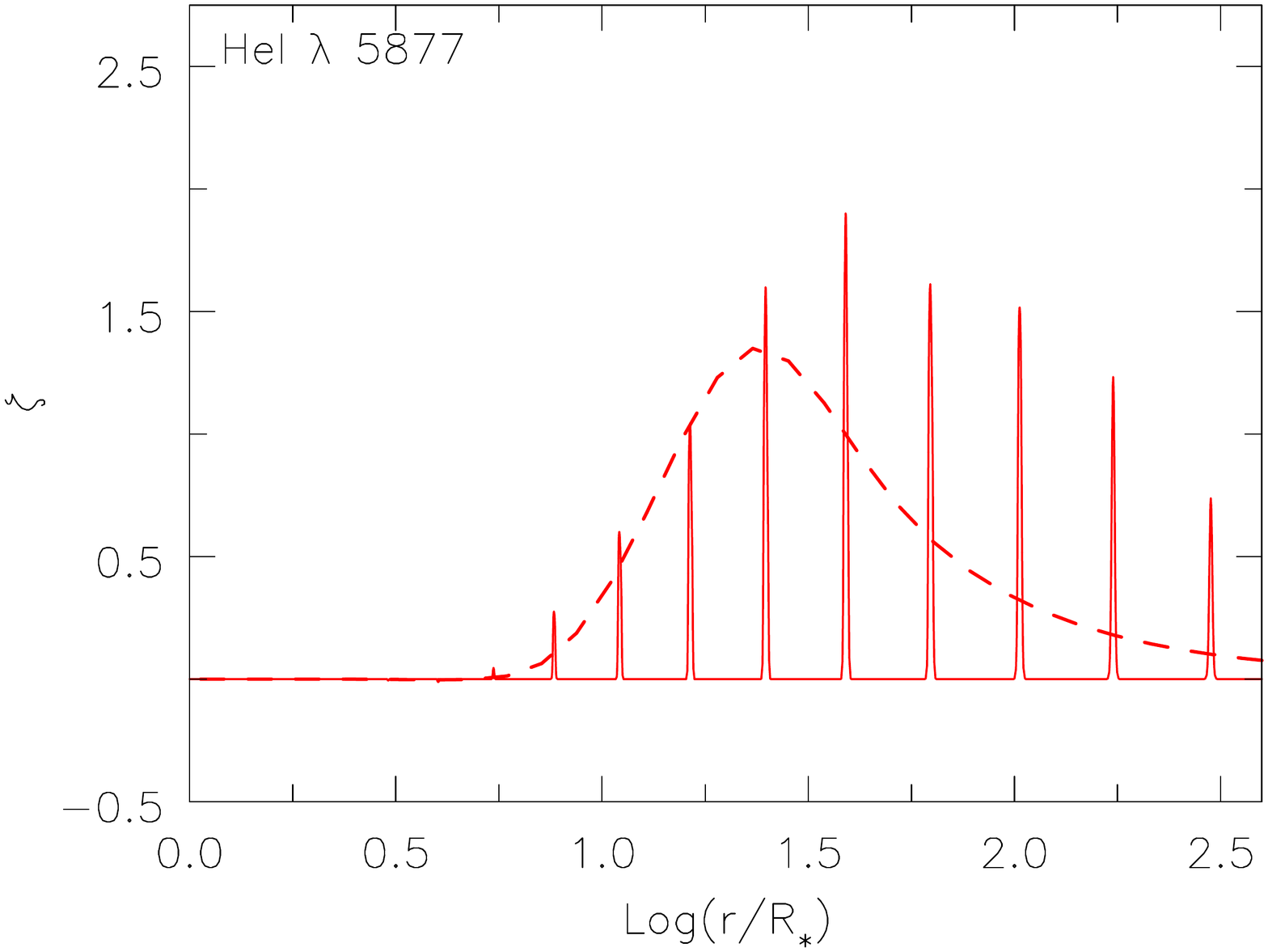}
		\includegraphics[scale=0.33,angle=-90,trim={2.5cm 1cm 0cm 2cm},clip]{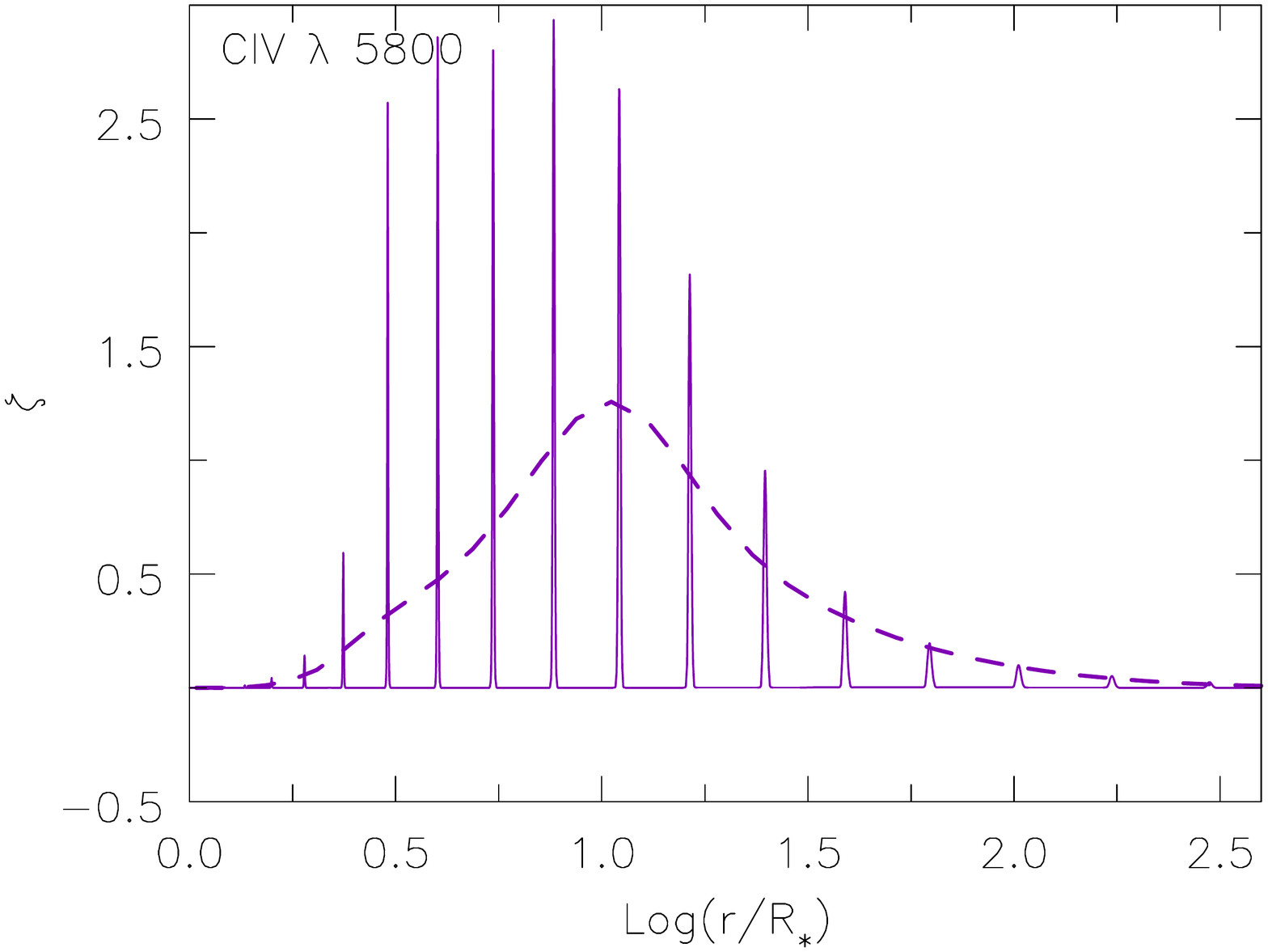}
		\includegraphics[scale=0.33,angle=-90,trim={2.5cm 1cm 0cm 2cm},clip]{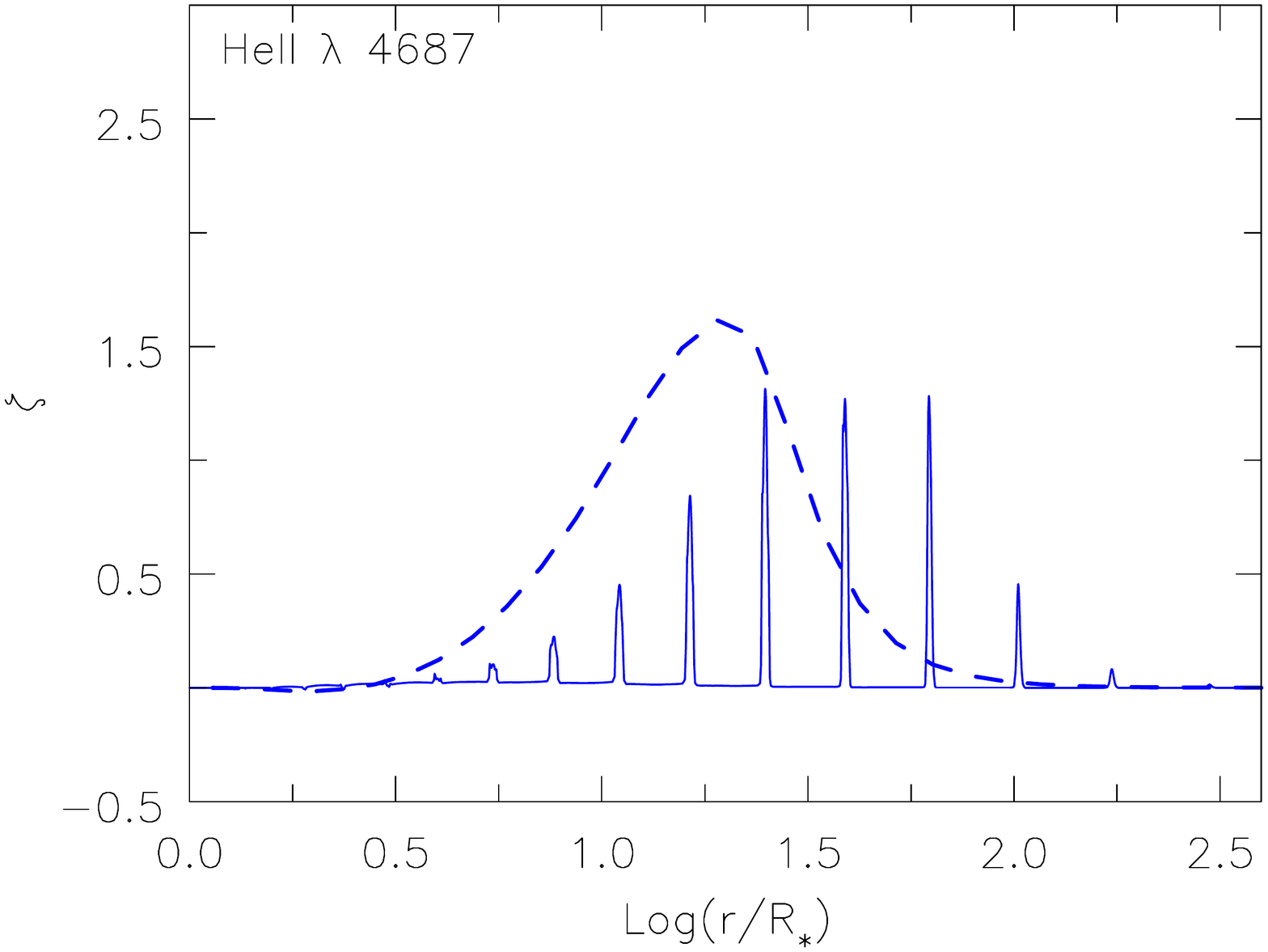}
		\includegraphics[scale=0.33,angle=-90,trim={2.5cm 1cm 0cm 2cm},clip]{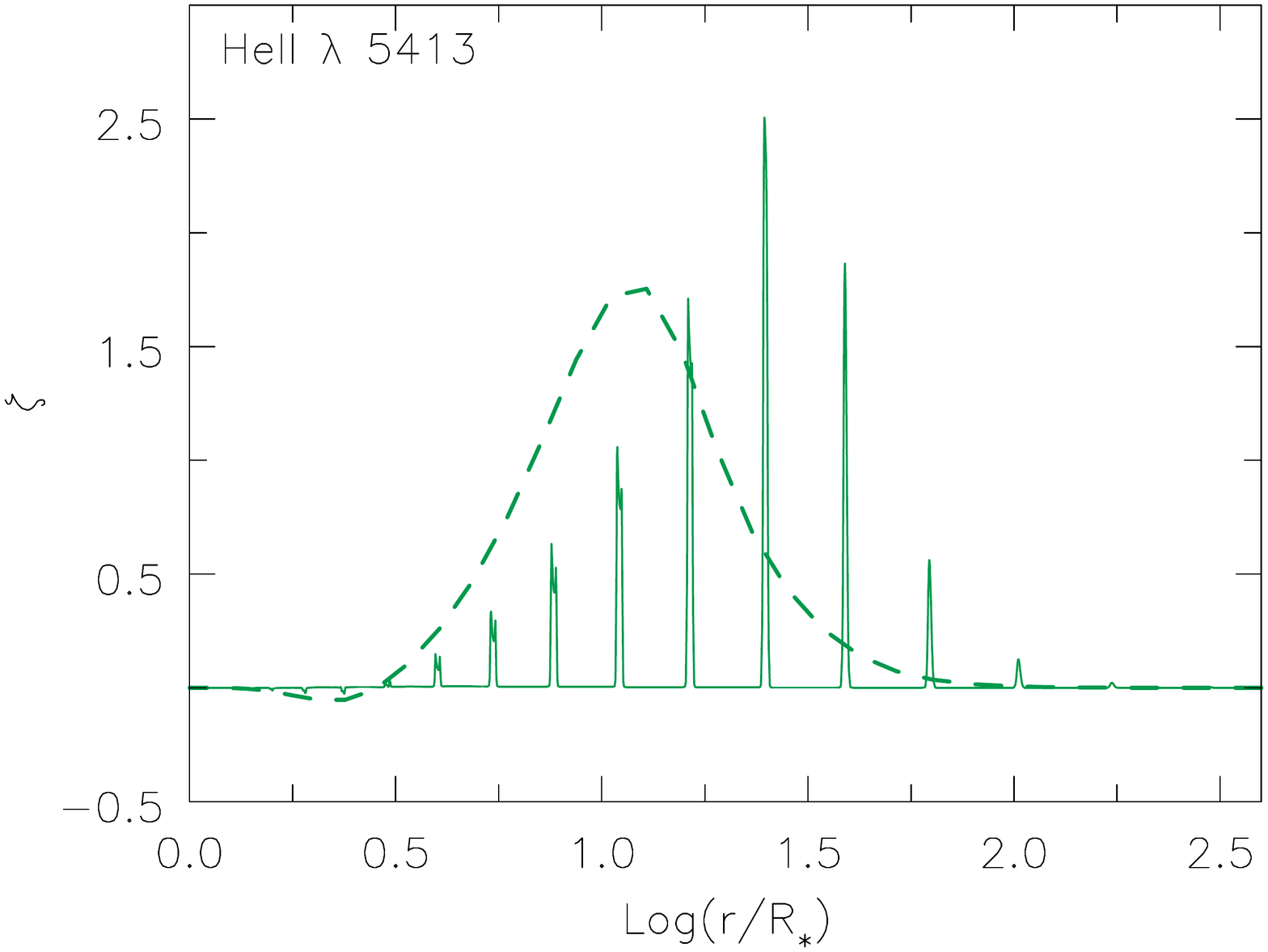}
		\caption{Illustration of the radial origin of \civdoubopt\ \ion{C}{IV}, \Heilineopt, and \Heiiline$\lambda$ 4687 and 5413 in the Shell \textit{(solid red)} and VFF \textit{(dashed blue)} models. The area under the dashed curve is unity while the area under the solid shell curves is 0.1. In the VFF model we see that lines originate over a wider range in the wind, while in the Shell model it is restricted to the shells. As discussed in the text, the shift outward in the formation zone of the \Heilineopt\ and \ion{He}{II} $\lambda\lambda$4687 and 5413 lines is due to higher optical depths in the Shell model. 
		}
		\label{fig_line_origin}
	\end{figure*}

	To aid future discussion we show in Fig. \ref{fig_line_origin} where the optical lines to be discussed originate\footnote{ As defined by \cite{Hillier1987}, the dimensionless quantity $\xi$ is given by $\alpha^{-1} N_u(r)\,r^3 \int^{+1}_{-1} \beta(\mu)\,\, \text{exp}[-\tau_c(\mu)]\, d\mu $ where $\beta(\mu)$ is the	angle dependent Sobolev escape probability, $\tau_c(\mu)$ is the continuum optical depth (absorption only), and $\alpha= \int_0^\infty \xi d\log(r/R_*)$ is the normalization constant which is proportional to the line flux. Thus the	area under the curve highlights the locations where the line emission originates. The integral is the angle-averaged escape probability corrected for the continuum opacity and $N_u$ is the population of the upper level. Assuming the Sobolev approximation holds, this quantity becomes large in the regions where the lines originate.}. Lines originate over a range of radii and different lines  originate in different regions. As expected, the line emission in the Shell model primarily originates in the shells. Somewhat surprising, however, is that the emission is shifted (on average) to somewhat larger radii\footnote{For the Shell model we also used the Sobolev approximation which has accuracy issues when treating clumps. However using an alternative technique (where we record the last interaction of a photon before escaping to the observer) shows a similar shift of the emission region to larger radii.}.

\subsubsection{Number of shells} \label{subsec_num_shells} 

\begin{figure}  
		\centering
		\includegraphics[height=\columnwidth,angle=-90,trim={2.5cm 1cm 0cm 2cm},clip
		]{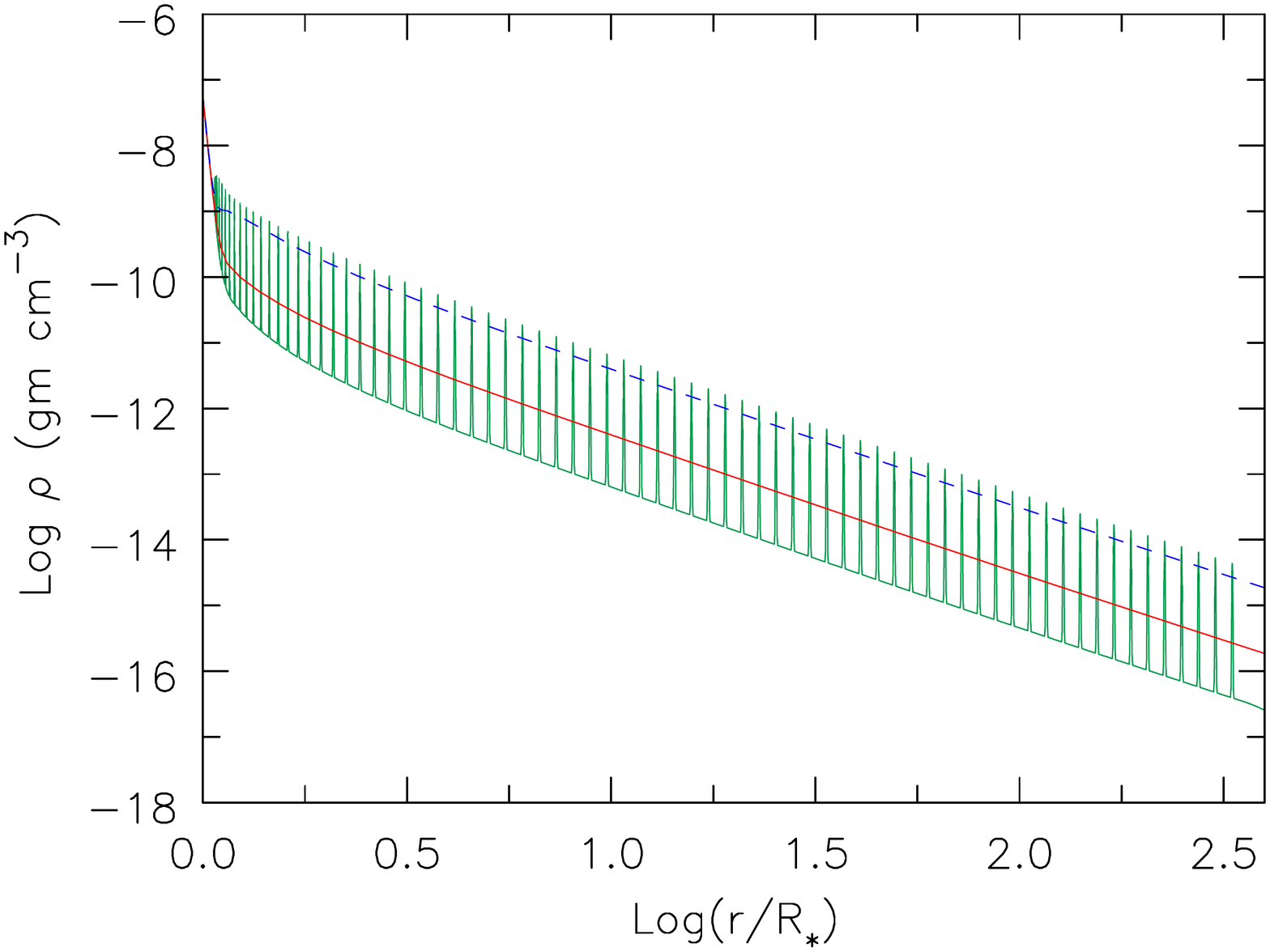}
		\caption{Same as Figure \ref{fig_1_dens}, but for a Shell model using thinner shell profiles. Using a thinner profile, while maintaining the same volume-filling factor of 0.1, resulted in a quadrupling of the number of shells (N$_{\rm Shell}$ = 79) in the wind.
		}
		\label{fig_den_4x_sh}
	\end{figure}

	One possible reason for the lack of flux at line centre of the He emission lines is that the shells we used were too thick. We therefore ran another Shell model in which we quadrupled the number of shells (to N$_{\rm Shell} =$ 79) while keeping the volume-filling factor fixed. Because the volume-filling factor is fixed, the shells are necessarily thinner. Figure \ref{fig_den_4x_sh} shows the density structure of the wind with the enhanced number of shells. 
	
\begin{figure}  
		\centering
		\includegraphics[height=\columnwidth,angle=-90,trim={2.5cm 1cm 0cm 2cm},clip]{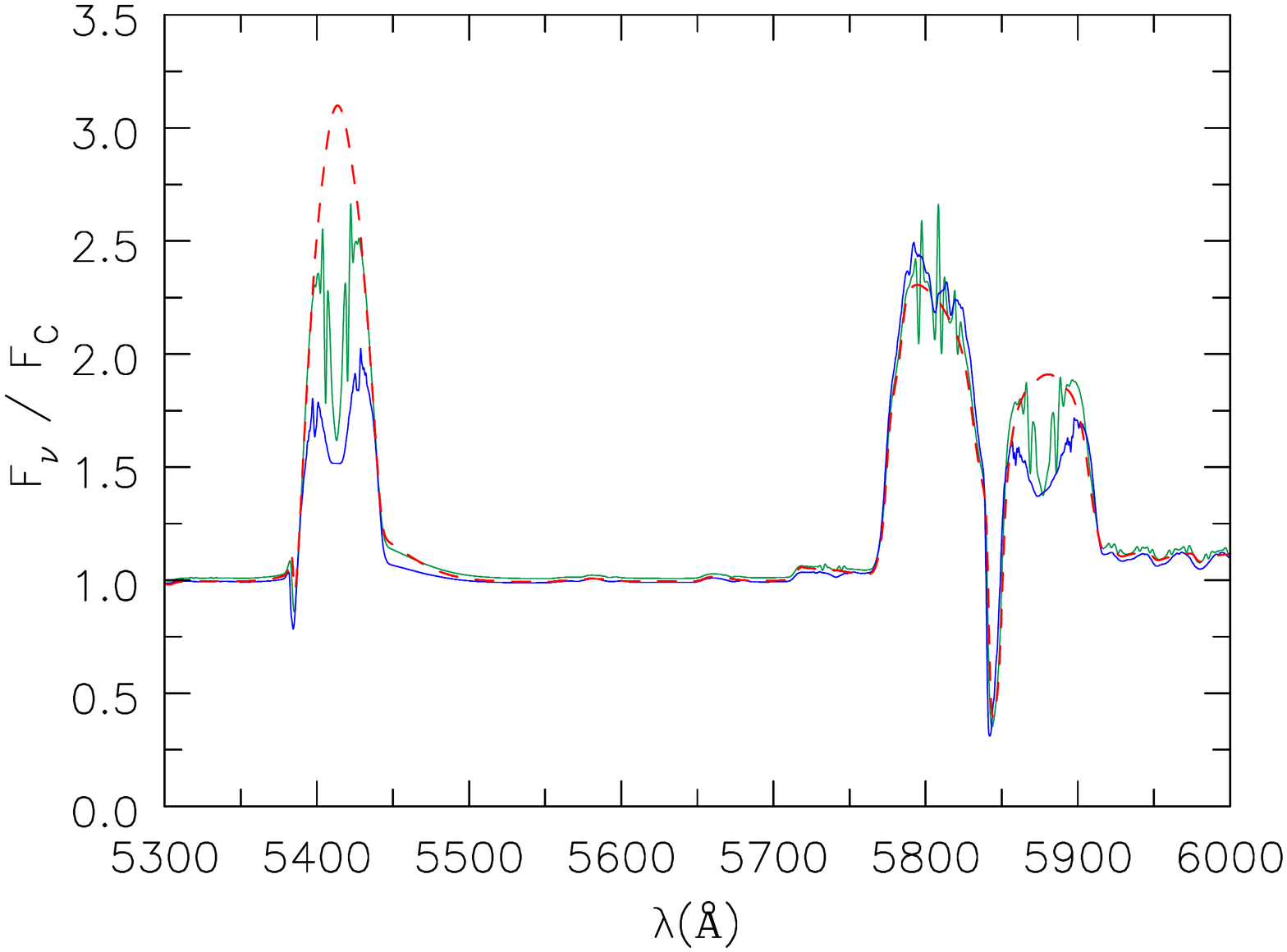}
		\caption{Plots of \Heiiline{5413}, \civdoubopt, and \Heilineopt\ are shown for a Shell model with N$_{\rm Shell} =$ 19 \textit{(blue solid curve)} and 79 \textit{(green solid curve)}. A VFF model \textit{(red dashed curve)} is also shown. All models use a microturbulent velocity of 50 \kms. While increasing the number of shells has improved the agreement between the Shell and VFF emission profiles of \Heiiline{5413} and \Heilineopt, it has not filled in the dip at line centre.
		}
		\label{fig_spec_4x_sh}
	\end{figure}

	In Fig. \ref{fig_spec_4x_sh}, we again plot  the line profiles for \Heiiline{5413}, \civdoubopt, and \Heilineopt\ for a Shell model using 19 and 79 shells, and compare it to the VFF model. While increasing the number of shells enhances the emission profile of \Heiiline{5413} and \Heilineopt, it does not fill in the dip at line centre.

	While increasing the number of shells improved the emission of \Heiiline{5413} and \Heilineopt, the general profile for \civdoubopt\ remained similar to that in the VFF model (see Fig. \ref{fig_spec_4x_sh}). This provides a clue as to why the \ion{He}{} lines are heavily reduced at line centre in the Shell model compared to the VFF model --  \civdoubopt\ are optically thin whereas the He lines (particularly \ion{He}{II}\ transitions) are susceptible to an optical depth effect, and this susceptibility may be enhanced when clumps are treated using shells. 

\section{Shell method versus the volume-filling factor approach}
     \label{shell_vs_vff}
	
\subsection{Basics}

    As apparent from Fig.~\ref{fig_1_dens} the density in a shell, and in the VFF model at the same radial location, are very similar. Thus, to first order, we would expect the optical depths to be similar. However, because of the assumptions built into the VFF model this need not be the case.

    For the following analysis let us assume the number densities for the two levels involved in a transition are identical in the two approaches. As a consequence, the Sobolev optical depth computed using those densities will be identical. However what is important is not the Sobolev optical depth, but the effective optical depth. Unlike the Sobolev optical depth, the effective optical depth will depend on the Doppler width\footnote{The Doppler width, $\Delta \nu$,  is defined by $\Delta \nu = \nu_o \varv_{\,\rm D+T}/c$ with $\varv_{\,\rm D+T} = (2kT/m_{\rm atom} + \varv_{\rm Turb}^2)^{1/2}$. In our analysis, we ignore the thermal term, which forces all atoms in our model to have the same Doppler width and Sobolev length (i.e. $L_{\rm Sob} = \varv_{\rm Turb}\times dz/d\varv$). The thermal Doppler width of He at 40,000\,K is $\sim$13\,\kms, which is generally much less than the adopted turbulent velocities. Microturbulent velocities of 20 to 100\,\kms are routinely adopted in spectral modelling of WR stars \citep[e.g.][]{Hamann2004}.} of the line.

    Let the line opacity be $\chi_L=\bar\chi \phi(\nu)$ where $\bar\chi$ will have units of Hz/cm. The classic Sobolev optical depth $\tau$ is then given by 
    $$\tau=  {\bar\chi c\over \nu_o} {dz \over dV}$$ 
    where $dV/dz$ is the velocity gradient along the direction of interest, $\nu_o$ is the line frequency, and $c$ is the speed of light. However, in the VFF approach we assume that the medium is clumped -- in practical terms we assume that there are ``many" clumps per Sobolev length (the length scale on which the velocity changes by a Doppler width). Multiplying the Sobolev length by the volume-filling factor gives the effective Sobolev optical depth, \taueff, defined by 
    $$\taueff = f_V {\bar\chi c \over \nu_o} {dz \over dV}\,\,$$
    and $\bar\chi$ is computed using the clump density. For populations that scale as the density squared, the Sobolev optical depth in an unclumped model, and the equivalent clumped model will be the same (ignoring ionization changes). This occurs because the mass-loss rate in the clumped model is scaled by a factor of $\sqrt{f_V}$. In the comoving-frame calculations the same result is achieved by multiplying the opacities and emissivities by $f_V$ before solving the radiative transfer equation.

    For the Shell model, we cannot simply multiply the Sobolev optical depth by a filling factor.  While our Shell model  has a well-defined volume-filling factor, the factor that modifies the Sobolev optical depth to yield an effective optical depth is also a function of the clump size, and further, will vary with position in the clump.
    
     We now discuss two particular cases that highlight the differences, and provide insights into the central absorption in our spectra computed with the Shell model.

\subsubsection{Line optical depth along a radial ray}	
    \label{subsubsec_line_tau_rad}

\begin{figure}  
		\centering
		\includegraphics[height=\columnwidth,angle=-90]{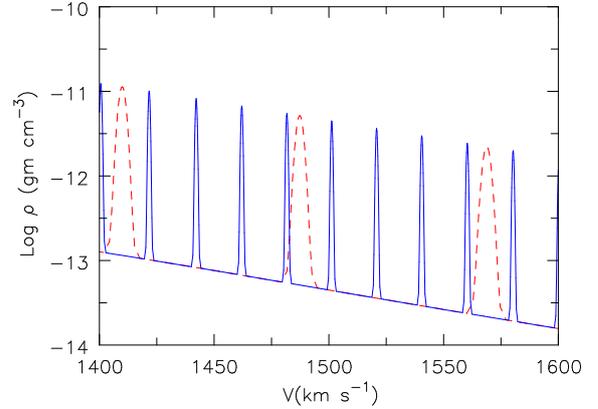}
		\caption{An illustration of the density profile for Shell models with $N_{\rm Shell}$ = 19 (\textit{red dashed curve}) and 79 (\textit{blue solid curve}) as a function of velocity. If the microturbulent velocity is small, a line photon will only be resonant in a single shell. However, a photon can interact with more shells if we increase the microturbulent velocity, or if we increase the number of shells (holding $f$ fixed).
		}
		\label{fig_den_vel_zoom}
	\end{figure}

    The number of shells across one Doppler width, for a fixed volume-filling factor,  will depend on the thickness of the shells. This is readily apparent from Fig. \ref{fig_den_vel_zoom} where  we plot a section of the density profile of the wind for two Shell models (one with 19 shells, the other with 79 shells) with the same  VFF of 0.1.  The number of shells within one Doppler width increases with the microturbulent velocity, and with the number of shells (i.e. decreasing shell width). If we assume a constant microturbulent velocity of 100 \kms, then across one Doppler width, a photon interacts with two shells for n=19, and with five shells for n=79. The use of a fixed Doppler width is obviously a gross simplification -- there is likely a wide range in velocities, and in the scale of velocity variations, and this will need to be explicitly taken into account in the solution of the transfer equation.
    
\subsubsection{Line optical depth at shells' intersection} 
	\label{subsec_tau_vturb}
	
    The lack of emission at line centre in the He lines suggests that we are receiving less emission from material with low projected velocities -- i.e. material in a (thickish) plane perpendicular to our line of sight and that passes through the centre of the star. We show below that this lack of emission arises from simple geometrical considerations, and optical	depth effects.

\begin{figure}  
		\centering
		\includegraphics[width=\columnwidth,trim={8cm 2.cm 8cm 2.5cm},clip]{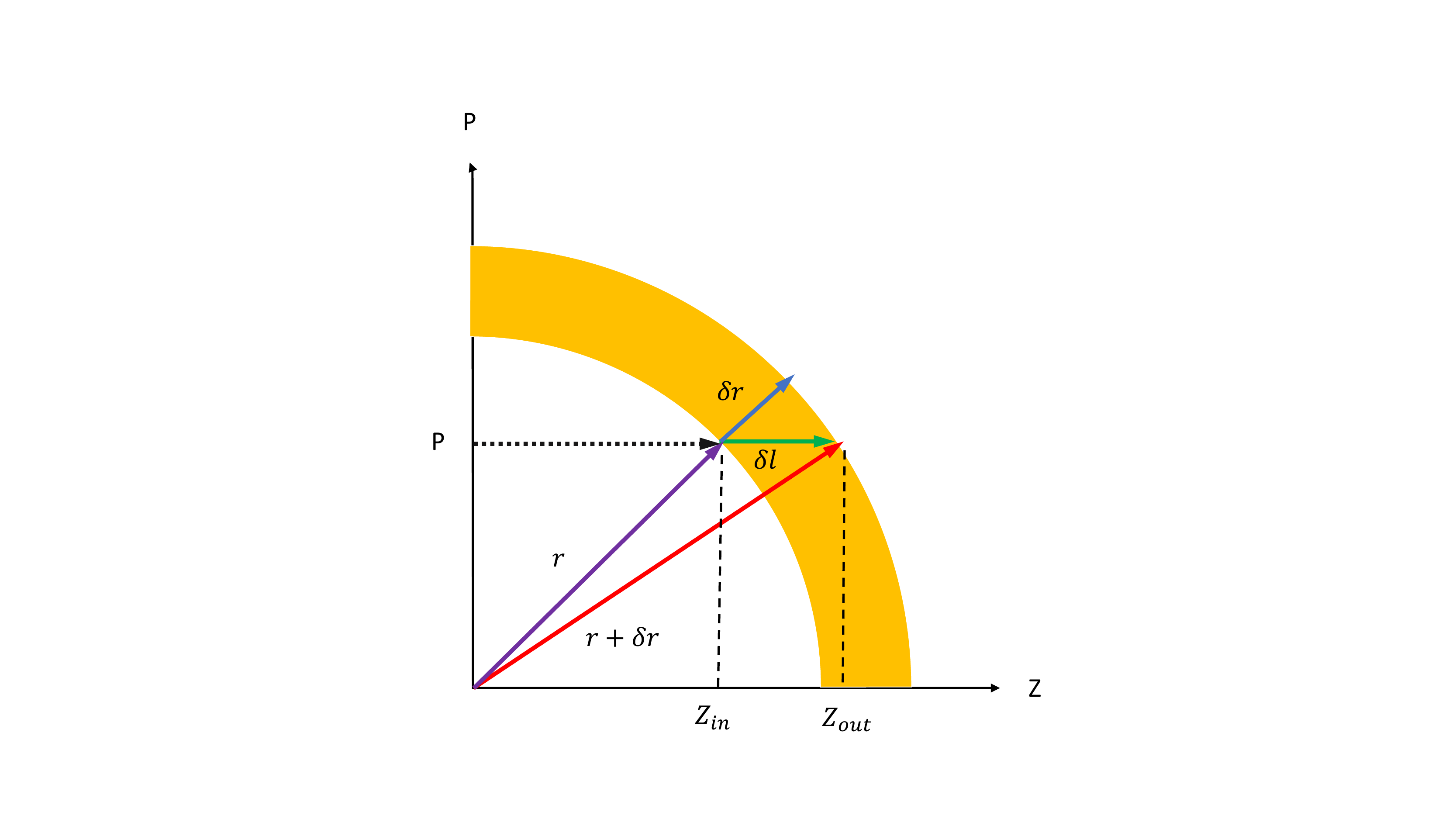}
		
		\caption{A simple sketch of a shell. $\delta r$ is the thickness of the shell, $\delta l$ is the path length of the shell at an impact parameter $p$, and all other symbols have their usual meanings. As we approach the shell limb, the path length increases. This well-known effect is, for example, responsible for limb brightening in planetary nebulae.}
		\label{fig_cross_sec_sh}
	\end{figure}

\begin{figure}  
		\begin{minipage}[t]{\linewidth}
			\centering
			\includegraphics[width=1.05\columnwidth]{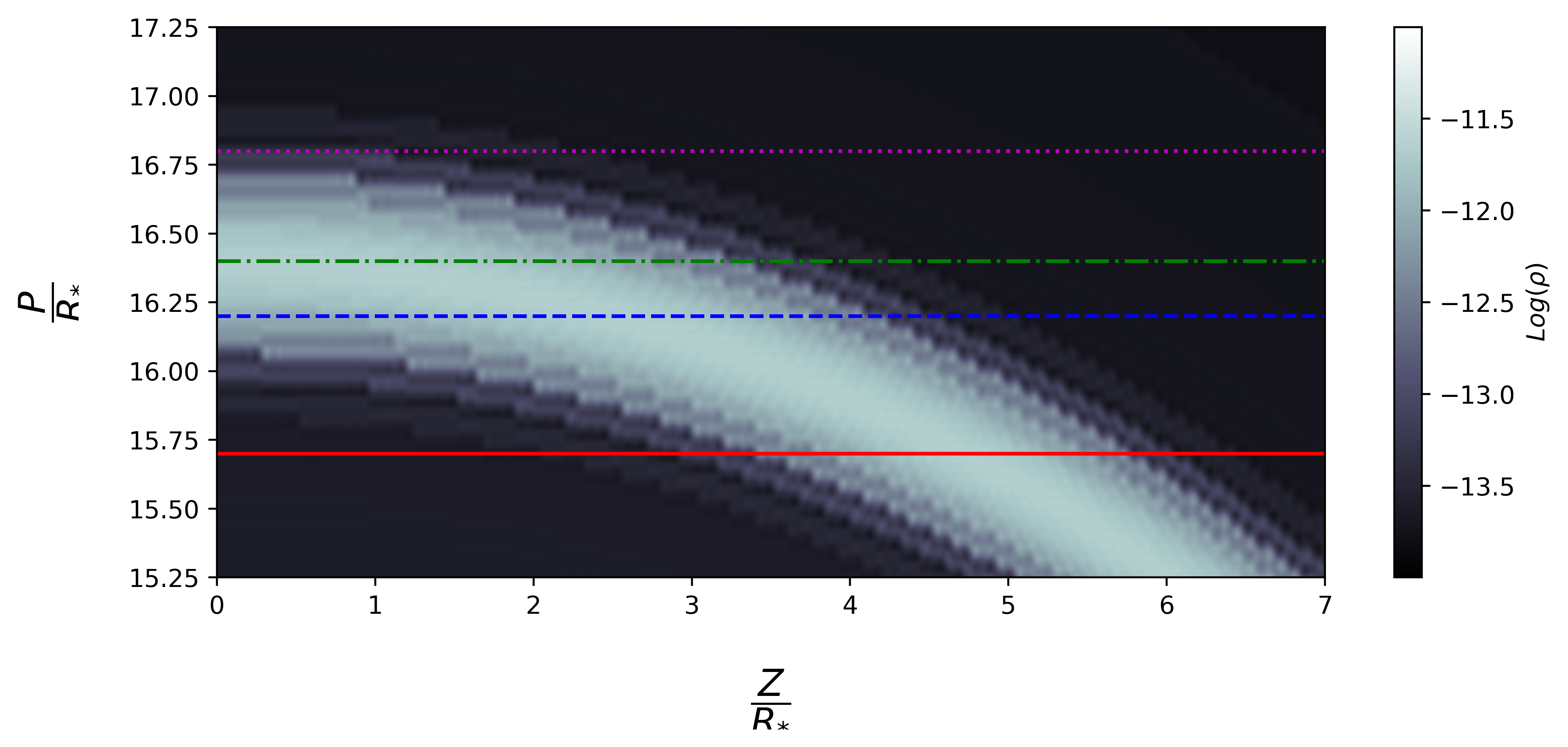}
			\includegraphics[height=\columnwidth,angle=-90,trim={2.5cm 0cm 0cm 0cm},clip]{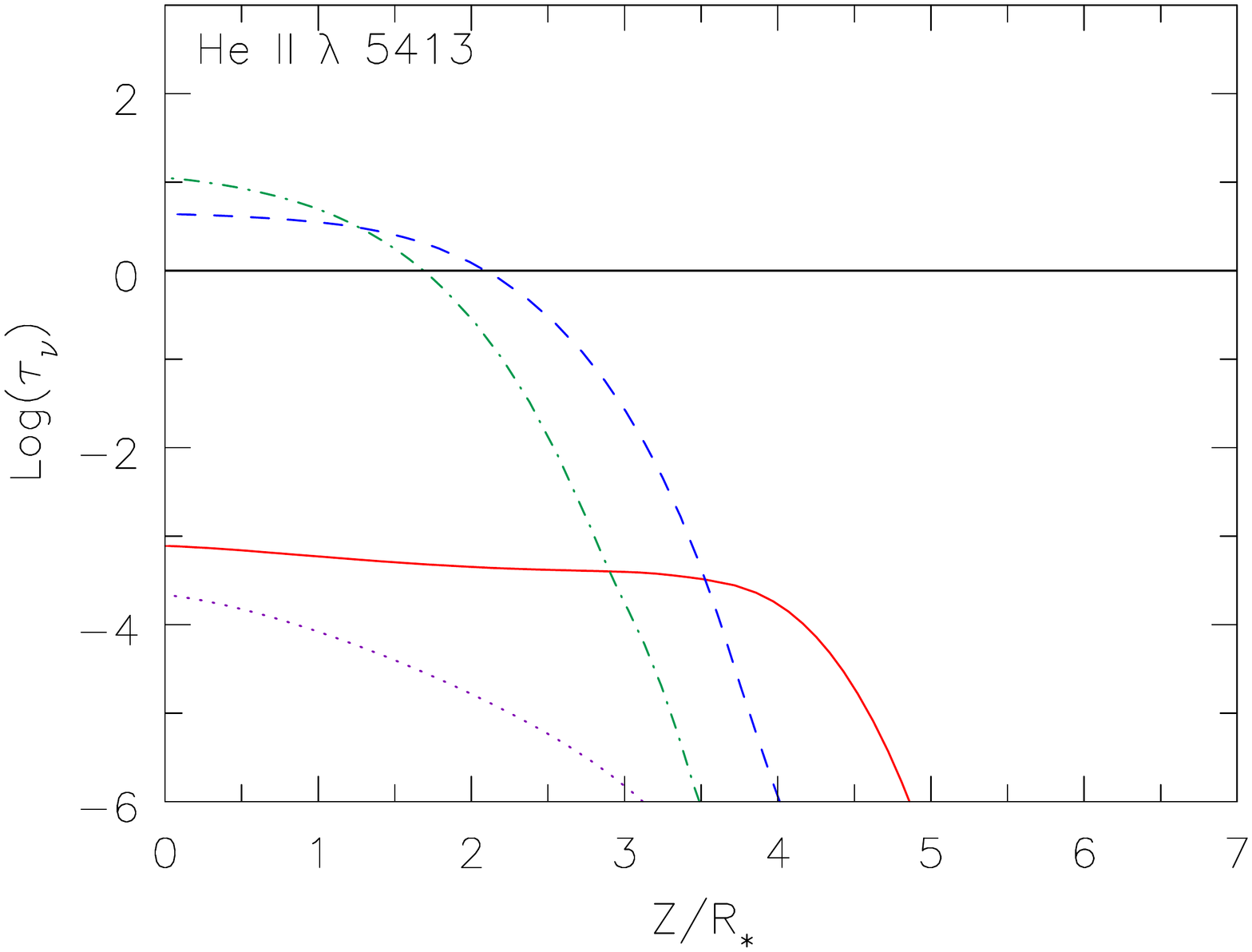}
			\includegraphics[height=\columnwidth,angle=-90,trim={2.5cm 0cm 0cm 0cm},clip]{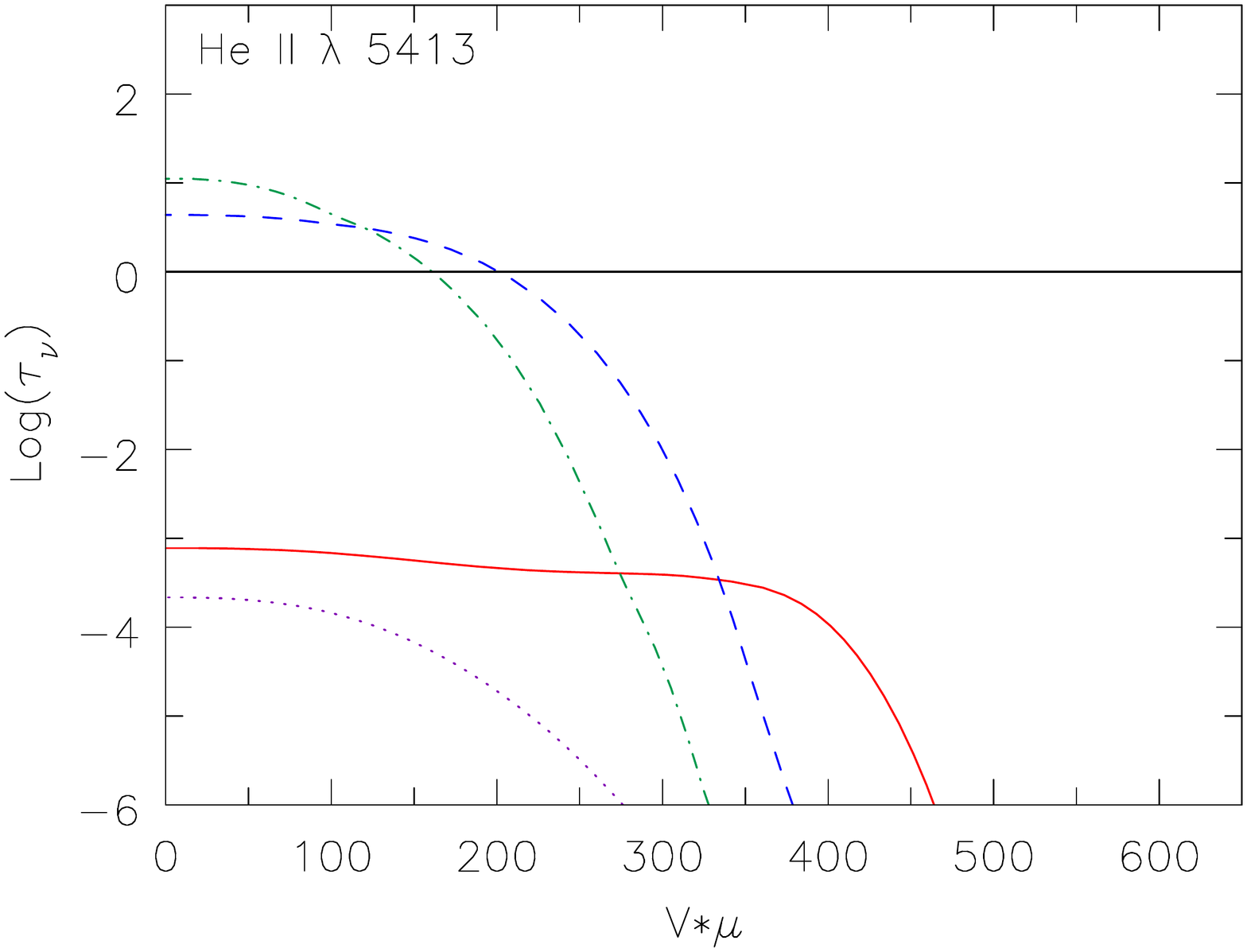}
		\end{minipage}
		\caption{The line optical depths for \Heiiline{5413} along the line-of-sight in $Z$ (\textit{Middle}) and projected velocity space (\textit{Bottom}) of several impact parameters through a shell located at r = 16.5 $R_*$. The dashed lines  correspond to the rays shown in the mass-density contour plot (\textit{Top}). The microturbulent velocity is set to 100\,\kms. Due to geometric considerations the optical depth varies strongly with the impact parameter. As we approach the shell limb, the increased path length leads to a larger optical depth. However, at the limb the variation in  optical depth is more complicated as it becomes sensitive to the precise location at which the ray intersects the shell.  The green curve (dot-dashed line) experiences a larger optical depth since it intersects the densest region of the shell).
		}
		\label{fig_taulip}
	\end{figure}

	A simple geometric relation between the shell's thickness, $\delta r$, and the path length, $\delta l$, can be derived using the sketch seen in Fig. \ref{fig_cross_sec_sh}. $\delta l$ is given by:	
	\begin{equation}
		\delta l = Z_{out} - Z_{in} = \sqrt{(r+\delta r)^2 - p^2} - \sqrt{r^2 - p^2}\,\,\, (\hbox{for}\,\, p \le r),
		\label{eq:sh_cs}
	\end{equation}
	where $r$ is the inner radius of the shell and $p$ is the impact parameter. As $p$ increases the path length through the shell increases, with $p = r$ yielding the largest path length: $\delta l \approx \sqrt{2 r \delta r}$. 	For $ r+\delta r > p> r $ we have $\delta l = \sqrt{(r+\delta r)^2 - p^2}$, and this also generally exceeds the radial thickness of the shell.  When $r = 16 R_*$ and $\delta r = 0.75 R_*$, as seen in Fig. \ref{fig_taulip}, $\delta l \approx 6.5 \delta r \approx 4.9 R_*$. Thus line emission originating from large impact parameters will experience a greater optical depth.

	To help illustrate this, Fig. \ref{fig_taulip} plots the line optical depth of \Heiiline{5413} at line centre for several impact parameters, striking near the limb of a shell located at r = 16 $R_*$. The radial velocity width of the shell is 10 \kms. A mass density contour plot is shown on top with several lines-of-sight intersecting the shell. These lines-of-sight yield different line optical depths, as shown in the lower two plots. 

	The optical depth increases rapidly once the shell is intersected (blue and green rays). The optical depth of the dot-dash line is larger since it intersects the densest region of the shell. Notice how the optical depth is significant over an extended region -- roughly 200\,\kms ($z/\Rstar=2)$ which corresponds to two Doppler widths.  As readily apparent, the escape of photons that will be observed at line centre are not influenced by the interclump medium -- mostly they will be absorbed before they reach the boundary of the clump. This is not true for rays with zero impact parameter, since the radial shell width is roughly one Doppler width. The solid curve is low since the	``shell" is not resonant for this impact parameter (i.e.~$\mu V \ne 0)$. Both the solid and dotted curves are only ``resonant" with material in	the interclump medium.
		
	In the present case we see that the effective optical depth for photons near line centre is simply ${\bar\chi c\over \nu_o} {dz \over dv}$ -- there is no reduction by the interclump medium.
	
\begin{figure}  
		\centering
		\includegraphics[height=\columnwidth,angle=-90,trim={2.5cm 0cm 0cm 0cm},clip]{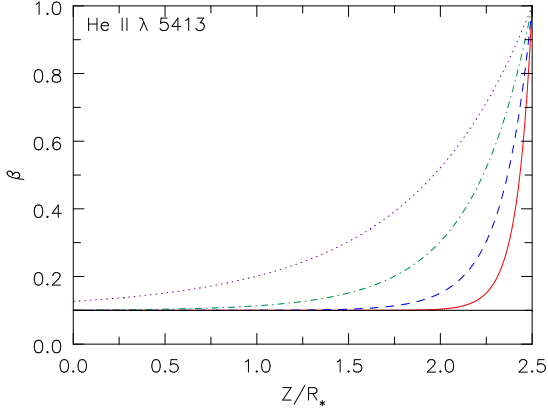}
		
		\caption{The escape probability for \Heiiline{5413}\ as a function of the Z-coordinate along the line-of-sight corresponding to the green, dashed line seen in Fig. \ref{fig_taulip}. We can represent the path through the shell with a slab of constant optical depth, $\tau = 10$, of a similar length, L = 2.5 $R_*$. The lines show the escape probability of \Heiiline{5413} with microturbulence of 25 (\text{red solid curve}), 50 (\text{blue dashed curve}), 100 (\text{green dash-dotted curve}), and 200 (\text{purple dotted curve}) \kms; the black line at 0.1 is the Sobolev escape probability. As expected, the escape probability increases with increased microturbulent velocity. 
		}
		\label{fig_esc_turb}
	\end{figure}

	An important parameter that influences the effective optical depth of the clump is the microturbulent (Doppler) velocity -- a larger value will enhance photon escape (from a single slab) since the slab width, expressed in Doppler widths, is smaller. This is illustrated in Fig.~\ref{fig_esc_turb} in the simple case of a photon escaping a homogeneous slab with constant opacity; this situation is a simplified representation of a photon traversing the densest path of the shell. In the VFF approach, the effective escape probability would be 0.63, much higher than the values shown in Fig.~\ref{fig_esc_turb}.

\begin{figure}  
		\centering
		\includegraphics[height=\columnwidth,angle=-90]{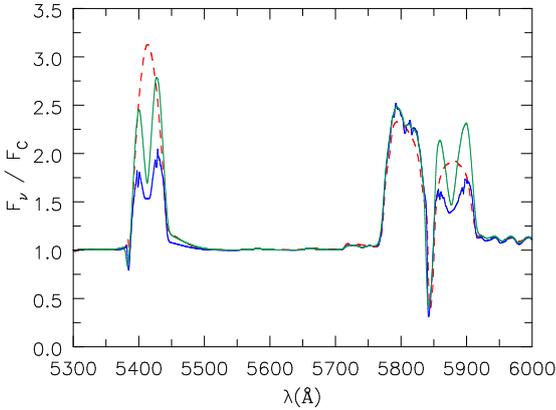}
		\caption{ \Heiiline{5413}, \civdoubopt, and \Heilineopt\ line profiles for a Shell model with a fixed (i.e. depth and species independent) microturbulent (Doppler) velocity of 50 \textit{(blue solid curve)} and 200 \kms \textit{(green solid curve)}. A VFF model \textit{(red dashed curve)}, using a microturbulent velocity of 50 \kms, is also shown.
		}
		\label{fig_spec_vturb}
	\end{figure}

	 In Fig. \ref{fig_spec_vturb}, we again plot \Heiiline{5413}, \civdoubopt, and \Heilineopt\ for a Shell model using a fixed microturbulent velocity of 50 and 200 \kms, and compare it to the VFF model; the VFF model uses a microturbulent velocity of 50 \kms. While increasing the microturbulent velocity from 50 to 200 \kms\ improves the quality of \Heiiline{5413} and \Heilineopt, the dip at line centre persists. This is unsurprising -- from Figure~\ref{fig_taulip} we see that the size of the clump is still comparable to the Doppler width.

\begin{figure}  
		\centering
		\includegraphics[width=\columnwidth]{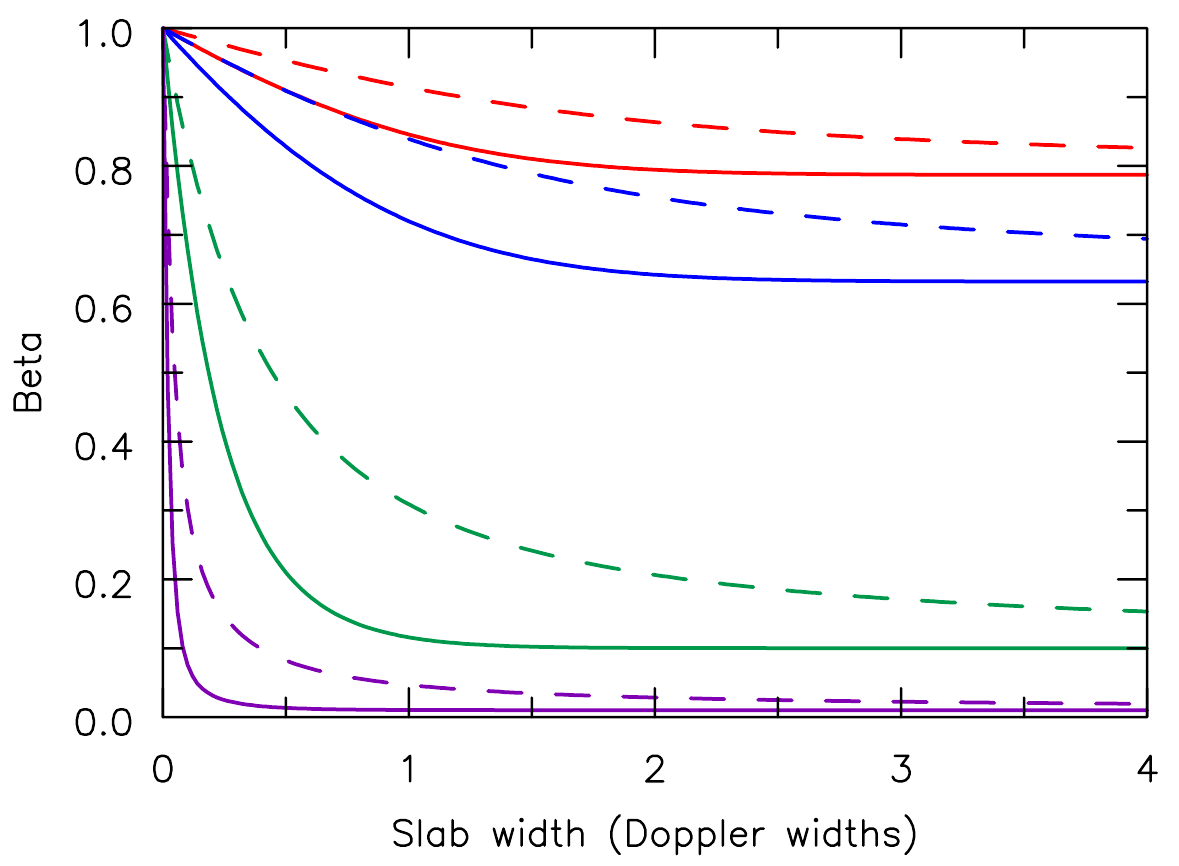}
		\caption{An illustration of the local radial escape probability as a function of location within the slab (averaged over the line profile) (\textit{solid curve}) and the spatially-averaged escape probability (\textit{dashed curve}) as a function of slab width. Each of the colors represents a different Sobolev optical depth -- 0.5 (\textit{red}), 1.0 (\textit{blue}), 10 (\textit{green}), and 100 (\textit{purple}). Once the distance from the slab edge exceeds a Doppler width, the local escape probability quickly approaches the Sobolev value. However the depth averaged escape probability is significantly higher, even when the clump width is a few Doppler widths, due to the enhancement in photon escape near the clump edge.
		}
		\label{fig_esc_slab}
	\end{figure}

    Changing the microturbulent velocity is similar to changing the intrinsic size of the clump (Figure ~\ref{fig_esc_slab}). With a larger microturbulent velocity (smaller clump) we approach the conditions necessary to satisfy the VFF approach. The two approaches are not identical -- when the microturbulent velocity is negligible, the Doppler width determines the Sobolev length, and this is species dependent, and consequently the ``effective'' size of the clump will be species dependent. However, once the microturbulent velocity exceeds that of hydrogen, the ``effective'' size of the clump will be independent of the atomic species.

\subsubsection{Broken shells} 
	\label{subsec_broken_sh}

	In our previous test we showed that a moderate increase in the
	number of shells did not fill in the dip at line centre for the optically thick \ion{He}{I} and \ion{He}{II} lines. Since the	dip stems from the emitted line photons travelling through a long path of an intact shell, we now test whether ``breaking'' up the shells can improve our spectra. By breaking up the shells, we reduce the line photon's path length through the dense shell, allowing more photons to escape through the interclump medium. 
	
	To synthesize a broken-shell spectrum, we used the 2D comoving frame/observer's frame code \formpol, a code that allows an accurate computation of observed spectra affected by polarization in stellar winds and supernova ejecta (Hillier and Dessart, in preparation). Ignoring the polarization aspect, the code is similar to another 2D code, {\sc obs\_2d\_frame} \citep{Busche2005}. It differs in that it can already handle the emissivity, opacity, and radiation field files that contained the 2D broken-shell inputs. With the given input, the code computes the emergent intensity along a set of rays that are specified by two polar co-ordinates --- the impact parameter and the polar angle $\delta$. A double integration of the intensities (with appropriate weights) over these two coordinates yields the final observed spectrum.
	
\begin{figure*}  
		\centering
		\includegraphics[width=10.5cm,trim={0cm 0.45cm 0.4cm 0.7cm}, clip]{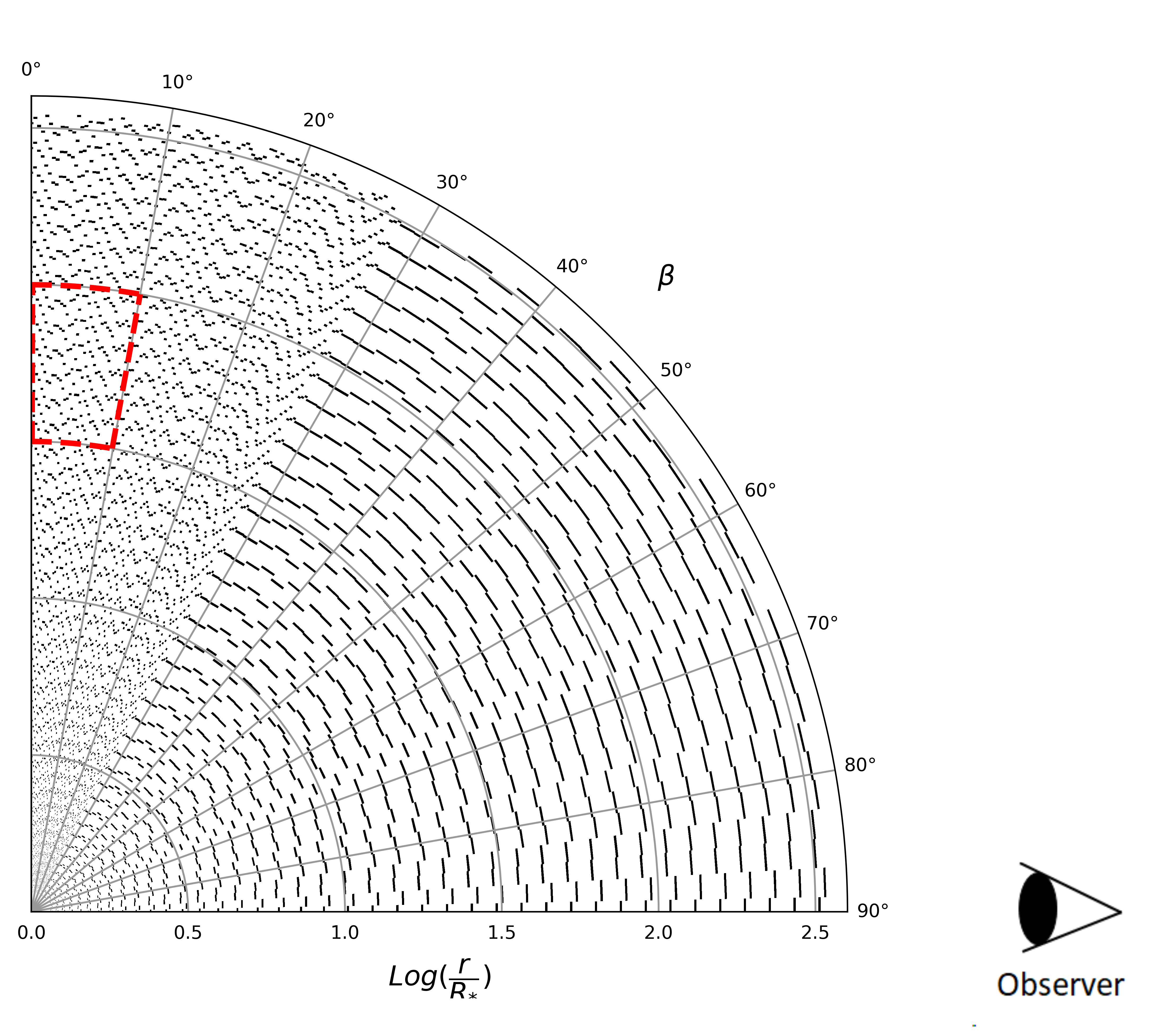}
		\centering
		\includegraphics[width=7.cm,trim={0.5cm 0cm 2cm 0cm}, clip]{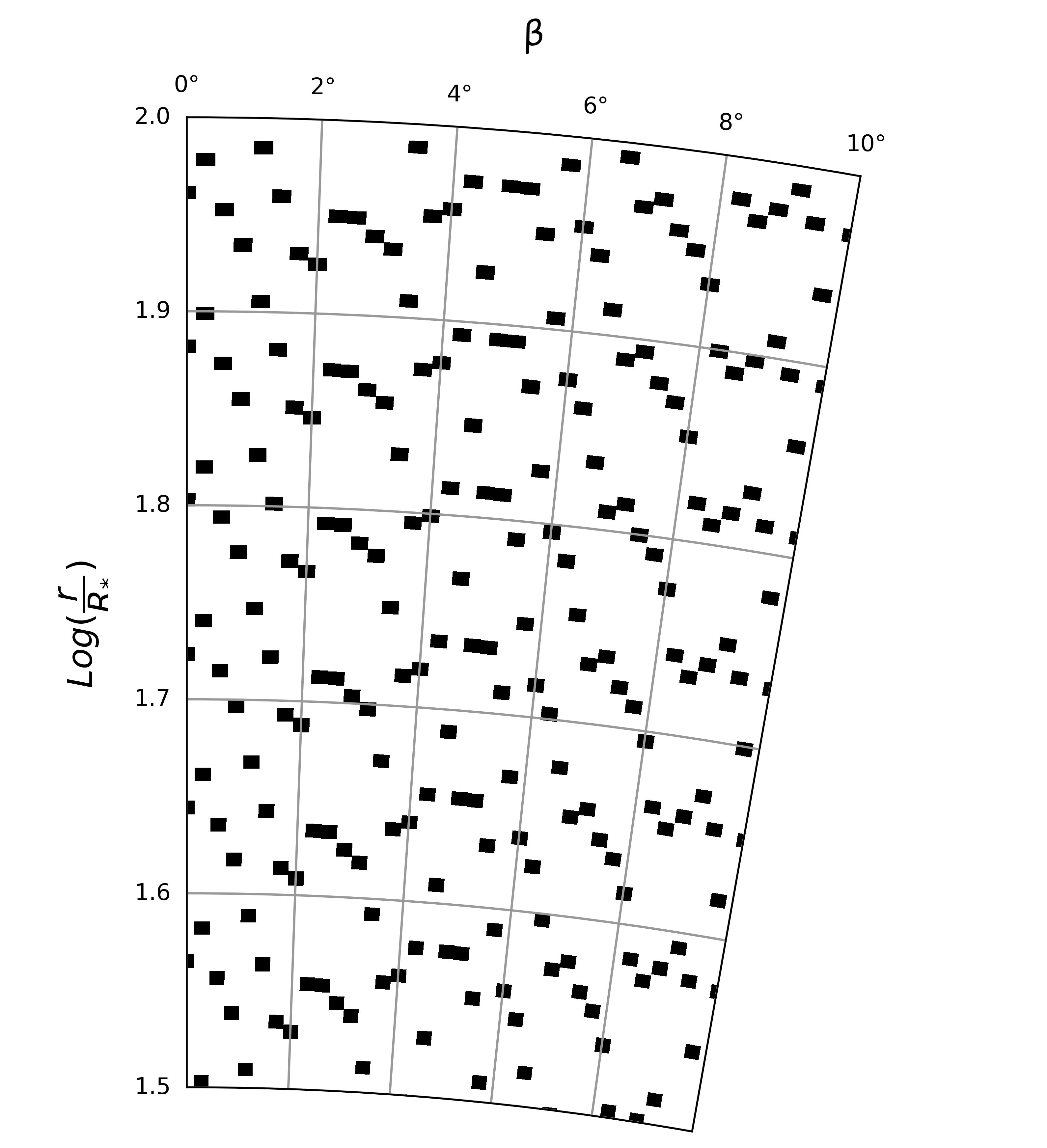} 
		\caption{The left panel shows the clump distribution used to synthesize Fig.~\ref{fig_beta_shift_spectra}. The observer is located in the equatorial plane on the right, and the figure shows the cross-section of the star viewed from the equatorial plane but rotated 90$^{\scriptstyle{o}}$ from the observer. The 44 shells have been randomly shifted radially by a factor proportional to the radial shell-width. The radial shift gives the shell a ``broken'' appearance, shown as the black features; the white background is the interclump medium. The shells are still continuous in the azimuth (i.e. about the vertical axis). The $\beta$ grid (with 129 angles) is finer towards the poles ($\beta = 0$) to allow for smaller broken shells in the wind. The right panel shows a close-up of the small broken shells around the pole (the wedge with the red dashed outline in the left panel). 
		}
		\label{fig_beta_shift}
	\end{figure*}

	The broken shells are constructed by reading in the opacities, emissivities, and radiation field of an unbroken Shell model. Using the density as a shell diagnostic, we determine the location of each shell and then apply a random radial shift (the shells retain their ordering) to each shell for each angle $\beta$.\footnote{ While $\beta$ and $\delta$ appear similar, they have fundamentally different meanings. $\beta$ is the polar angle in 3D. In Fig. \ref{fig_beta_shift} the observer is located in the equatorial plane on the right, and the figure shows the cross-section of the star viewed from the equatorial plane but rotated 90$^{\scriptstyle{o}}$ from the observer. $\delta$ is the 2D polar angle, and is measured on the star's disc, as seen by the observer.} Because the shells are not uniformly spaced, it is not feasible to uniformly shift them in $\log r$. The opacities, emissivities (including the electron scattering emissivity) are moved with the shell.  To ensure that the shells are sufficiently small such that it is optically thin at line centre for \Heiiline{5413} and \Heilineopt, a finer $\beta$ grid is used near the poles. Shells located in $\beta> 30^\circ$ are coarsely broken to save computational time, and because their path lengths are already comparable to the shell's thickness. Our final broken Shell model is shown in Fig. \ref{fig_beta_shift}, where we used 129 $\beta$ points. The black features are the broken shell and the white background is the subtracted interclump medium. It should be noted that while the shells are broken in the polar angle direction, they are still coherent in the azimuth direction.
	
	The lateral length ($\delta s$) subtended by the broken clump is $\approx\delta\beta\,\times r$, where $r$ is the clump's radial position and $\delta\beta$ is the angle subtended by the clump. In the model presented in Fig. \ref{fig_beta_shift}, $\delta\beta$ is about $0.3^{\scriptstyle{\rm o}}$ near the poles and thus the lateral length is $0.005\, r$. This is about a factor of 2 less than the radial size of the clump (full-width at half-maximum).  Since the clump's radial thickness is proportional to $r$, this is true for all clumps.
	
	As we are only shifting the shells as a function of $\beta$, the shell structure in the equatorial plane is unaffected  (i.e. the shell will be continuous in azimuth). However, since we are only performing a formal integral, the spectrum for $\delta=0$ is equivalent to a Shell model in which the shells were broken in all directions. Conversely, the spectrum for $\delta=90$ will be	equivalent to a model with unbroken shells.\footnote{With this approach we are assuming that although the shells are randomly fragmented in $\beta$, the model, on average, is still spherically symmetric. Thus, we only need to compute the spectrum for one $\delta$ angle. With the choice $\delta=0$, we are effectively choosing a model in which the shells are randomly fragmented in all directions.}
	
\begin{figure*}  
		\centering
		\includegraphics[scale=0.65,trim={0.5cm 0.5cm 0.5cm 12cm},clip]{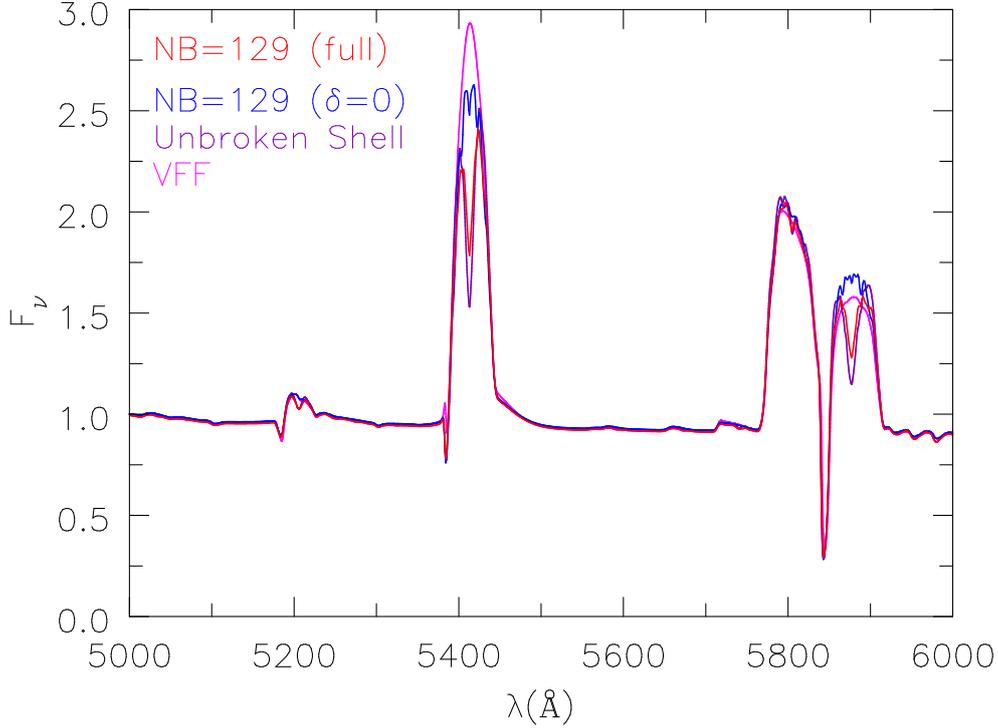}
		\caption{Model spectra for the region containing \Heiiline{5413}, \civdoubopt, and \Heilineopt\ computed using the broken Shell model. Also shown are the VFF model and an unbroken Shell model. In the broken shell spectra, the angle $\delta$ refers to the polar coordinate (see text). All models used a fixed microturbulent velocity of 100 \kms; the number of $\beta$ points (NB) used in the broken Shell model is 129. For the broken Shell model with $\delta=0$ the absorption dip at line centre in the He line profiles is much weaker, or has disappeared entirely.  When $\delta=90$ the broken Shell model's spectrum is indistinguishable from that of the unbroken Shell model, and for clarity is not shown. The spectrum computed using a full integration over the disc of the star is labeled ``full", and is similar to the unbroken shell spectrum because the shells are not broken in azimuth.
		}
		\label{fig_beta_shift_spectra}
	\end{figure*}

	In Fig. \ref{fig_beta_shift_spectra}, we show a spectral comparison of a Broken Shell model with the VFF and unbroken Shell models. All models use a fixed microturbulent velocity of 100 \kms, and the final number of $\beta$ points used in the broken Shell model is 129. We see that the emission profile's dip in the broken Shell model is filled in when $\delta=0$. Conversely, when $\delta=90$, the broken Shell model's spectrum is indistinguishable from the unbroken Shell model. 
	 
	These results confirm our postulate that the shells in our Shell models have enhanced optical thickness in the lines due to the long shell-path length seen by rays that are nearly tangential to the shell. The enhanced optical depth in the Shell model inhibits escape at line centre in the observer's frame, producing a central dip. Note that this effect will happen when performing radiative transfer calculations using the results of a 1D radiation-hydrodynamics model.

\section{Escape probability from multiple clumps}
    \label{sec_esc_mc}
    Previously we discussed the escape probability from a shell, and showed how the lateral extent of the shell in the Shell model influenced photon escape. Here we address the more fundamental question -- under what limits does the VFF approach hold? To do this we consider a constant volume-filling factor and change the width of the shells. For simplicity we consider only escape in one direction, and thus we do need to worry about the geometry of the shell.

    If we express the width of the shell in units of the Doppler width there are only three parameters that need to be considered: the volume-filling factor, the Sobolev line optical depth, and the shell width.

\begin{figure}  
		\centering
		\includegraphics[scale=0.85]{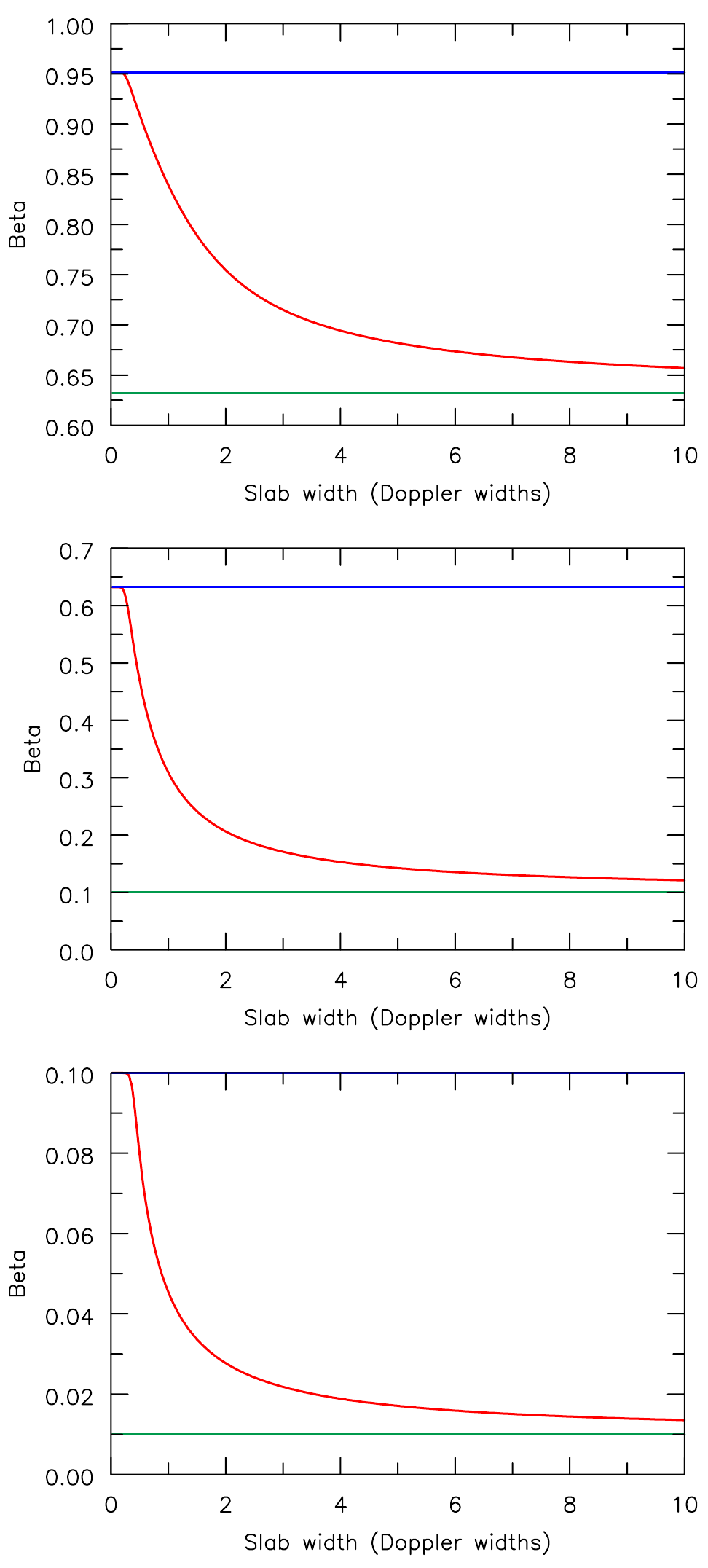}
		\caption{The net radial escape probability from a clump as a function of clump width measured in Doppler widths. A fixed velocity gradient and a constant volume-filling factor of 0.1 are assumed. We characterize the clump by  the local Sobolev optical depth. Three cases are shown: $\tau=1$ (top panel), 10 (middle), and 100 (bottom). When the clump is small (i.e. < 0.2 Doppler widths), the escape probability (red) approaches the escape probability for a homogeneous medium with an average density of $f$ times the clump density (blue).  As the clump size increases, the escape probability falls, and asymptotically approaches the escape probability for a uniform medium (pale green/dashed). As expected we recover the VFF  approach for small clump sizes ((i.e. < 0.2 Doppler widths) but for larger clump sizes the deviations become significant,
		}
		\label{fig_esc_sh}
	\end{figure}

    In Figure~\ref{fig_esc_sh} we show the net (i.e. allowing for optical depth effects in the emitting clump and additional clumps through which the photon travels) radial escape probability for three optical depths: 1, 10 and 100. For each optical depth we show two further lines. One is the Sobolev escape probability. The second line is the Sobolev escape probability if the VFF aproach is valid -- in this case $$\tau^{\scriptstyle VFF}_{\scriptstyle SOB}= f_V \times \tau^{\scriptstyle smooth}_{SOB}\,\,$$

    From the figure we see that the average escape proability from the shell only approaches the VFF result when the clump width is less than one Sobolev length, and the VFF approach will only be accurately valid when the shell width is less than about 0.2 Sobolev lengths, especially for high optical depths. The VFF approach will also only be valid when the velocity is smooth and monotonically increasing in $r$.

    As we increase the shell width, the escape probability approaches $\betasm$ very slowly. This occurs because photons at the edge of the clump can always escape more easily than predicted by the Sobolev optical depth. Let \betasob\ be the Sobolev escape probability from the clump, and let \betaed\ be the enhancement in escape probability due to the edge of the clump. Once the clump width is greater than a few Sobolev lengths, \betaed\ will be constant. Hence $$\langle \beta \rangle\, \approx\, (n \betasob + \betaed)/n$$ where $n$ is the width of the slab in units of Sobolev lengths, and thus the enhancement of \betaav\ over \betasob\ only falls by a factor of 2 as we increase the clump width from 5 to 10 Sobolev lengths.

    It is clear that the VFF approach makes rather strict assumptions about the clump size. In practice, however, the influence of larger clumps on spectral formation will be complicated. First, most lines form around $\tau=1$ and the requirements are somewhat less strict than for higher optical depth lines. If LTE holds (and the temperature is assumed fixed) errors in $\beta$ will directly influence the observed spectrum. However, when non~LTE holds the effect may be smaller, since the level population is itself influenced by the escape probability. If a line is the dominant depopulation mechanism for a level the strength will not necessarily change -- a halving of \betaav\ will simply lead to a doubling of the level population.

\section{Discussion}\label{sec_discuss}
	
	
	The volume-filling factor (VFF) approach is highly successful when used to treat clumpy winds in massive stars \citep{CrowtherO2002, Puls2006}, but the underlying assumption (i.e. a wind comprised of a homogeneous distribution of small-scale clumps embedded in a void background) was made for ease of calculation.  Additionally, when accounting for the influence of the wind's velocity field on line transfer, which reduces the size of the line interaction region to $\sim rV_{\rm Dop}/V(r)$, we would expect the VFF to be more accurate for the continuum (with a much larger scale length) than for lines. Thus it is somewhat surprising how well models that use the VFF approach replicate observed spectra of WR stars \citep[e.g.][]{Hillier1999, Hamann1998, CrowtherWR2002, Aadland2021, Aadland2022_arXiv}. 
	
	One might have expected that using shells as a method to treat clumping would be preferable given that it makes no inherent assumptions about the size of the clumps, and allows optical depths to be calculated from the supplied density and velocity profiles. However, our models revealed that the VFF and Shell models yield different  optical spectra, which has important implications for understanding clumping in WR winds.
	
	Our Shell model shows that the optically thick \Heiiline{5413} and \Heilineopt\ lines (the latter are normally optically thin) are heavily affected by the size and geometry of the shells along the plane perpendicular to our line of sight. In order to fill in the dip at line centre, our model required the shells to be broken into fine  fragments, as shown in Fig. \ref{fig_beta_shift_spectra}. This effect can be understood as follows.
	
	In the VFF approach, all wind material is found within the clumps, such that the opacity is enhanced by a factor of $1/f_{V}^2$ (for levels whose population is proportional	to the square of the density). However, as the clumps only occupy a volume $f(v)$, we scale this opacity by $f_{V}$. As such, the optical depth of the line is given by $\tau_{\rm line} = \int (\chi/f_{V})^2 \, (f_{V} \,dz)$, and hence scales as ${1/f_V}$\footnote{When we introduce clumping we scale the mass-loss rate by $\sqrt{f_V}$, and since the opacity is proportional to $\dot M^2$, the line optical depth is preserved.}. In the above, $\chi$ refers to the unclumped opacity.
	
	For the Shell model the opacity is also enhanced, in the shells, by $1/f_{V}^2$. However these shells are roughly uniform in density. If the shells are broad (i.e. many Doppler widths) there is very little reduction in opacity due to the interclump medium (the reduction that does occur arises from the clump edges), and hence the line optical depths are intrinsically higher than in the VFF approach. In the case of shells, the spatial extent varies with direction -- it is much larger for rays striking near the edge of the shells, and this is what causes the central reversal in the \ion{He}{II}\ profiles. When we break the spherical shells (as seen in Fig. \ref{fig_beta_shift}) our models more closely fulfill the conditions assumed by the VFF approach, and the spectra become more similar. 
	
	In principle the {\sc cmfgen} non~LTE calculations should also use fragmented shells. However, this will produce  a smaller change in line profiles  as line profiles are generally more sensitive to radiative transfer effects than how the level populations are computed  (as demonstrated by the wide use of the SEI [Sobolev with Exact Integration] approach \citep{Lamers1987} to understand	P~Cygni profiles in massive stars \citep{Groenewegen1989, Bianchi1994, Eenens1994, Zickgraf1996}).

	The results from our simulations indicate that clumps in HD~50896 are required to be small (i.e. less than a Sobolev	length) and numerous, regardless of the clump's optical thickness in a line. What matters most in determining the escape of a line photon, for a given volume-filling factor, is the size of the clump relative to the (directional dependent) Sobolev length.  The latter depends on the velocity gradient, the atomic mass, and the adopted microturbulent velocity.

	We conjecture that the observed line-profile morphology in some WNE stars might be a consequence of the size and shape of the clumps in their winds. In some stars the \ion{He}{II}\ line profiles are rounded (parabolic-like) while in others they are more Gaussian (and can be reproduced by theoretical models). \cite{Shenar2014} suggest that the rounded lines may result from the WN star having a strong magnetic field ($B_{\tau=2/3} \sim 20$ kG) which enforces co-rotation up to an inferred radius. Though we did not set forth to model these rounded line profiles in WR winds, it might be possible to adjust the distribution of small and large clumps in our broken-Shell model such that the filled-in line profiles look rounded without the need to invoke magnetic fields. However, the mechanism that gives rise to such a distribution of clumps requires further consideration.

    The very early Shell models of \cite{Hillier1991} did not find a central reversal in the \ion{He}{II} profiles. This is due to the weak clumping assumed in these models -- the effective volume-filling factor was 0.5 to 0.6. As a consequence, the enhancement in the optical depth within a clump is much smaller.

	
\section{Conclusions}\label{sec_conclusion}
	
	
	We have presented a comparison study of two procedures, our Shell approach, and the commonly adopted VFF approach, for modelling a dense wind such as that inferred for the WN4 star, HD~50896.  
	
	Unlike our results for the  star AzV83, we find that the VFF
	and Shell approaches do not yield consistent results. In the Shell approach, optically thick \ion{He}{II} lines in the optical are weaker than their VFF counterparts, and show a central absorption dip. The contrasting result with AzV83 most likely arises from two effects. First, and as apparent in this paper, different lines, with different opacities and distinct formation mechanisms, show distinct behaviours. Second, the wind of HD 50896 is much denser than that of AzV83.
	
	To understand our results we primarily focused on a few lines found in the optical region: \Heilineopt, \Heiiline{4687}, \Heiiline{5411}, and \civdoubopt. This set of lines encompasses several fundamental line-forming mechanisms that are at play in WR winds (i.e. collisional excitation and deexcitation, photoionization and recombination, line and continuum pumping). We have shown that the  helium ionization structure in the Shell models closely matches the radial trend seen in a VFF model -- the dominant ionization state of helium shifts from He$^{++}$ to He$^+$ at approximately the same radius. The DCs of the levels involved in the formation of the said lines show more complexity in the Shell models than in the VFF models, however the general radial trend seen in the VFF model is also seen for the dense shells. As a consequence of the lower density, the DCs in the intershell medium show much larger differences from the VFF model. However, for the lines discussed in this paper the intershell medium does not strongly affect the observed emission. This is not always the case -- in O stars, for example, the profile of the \ion{O}{vi}\ resonance doublet is strongly influenced by the interclump medium \citep{Zsargo2008,Flores2021}.
	
	Model spectra of HD~50896 computed using the VFF approach were in broad agreement with observations -- differences can be explained by the simplicity of the model and the choice of stellar parameters. However, when modelling spectra using the Shell approach, the \ion{He}{I} and \ion{He}{II} emission lines are noticeably weaker compared to the VFF models -- a ``dip'' was formed at line centre. The \ion{C}{IV} lines show relatively minor differences. We determined that the central depression arose from an optical depth effect. Line photons are less likely to escape the shell at low projected velocities, in part due to the photon traversing a large geometric path through the shell. Photons traveling in the radial direction are less influenced, although this depends on the adopted microturbulence. 
	
	Using a 2D auxiliary code \textsc{form\_pol} to compute the observed spectrum, the optical depth effect along the poles was resolved by ``breaking-up'' the lateral side of the shells into small and numerous fragments, allowing more photons to escape via the low-density intershell medium. This effect occurred regardless of the optical thickness of the clumps. It does not occur in the VFF approach since it explicitly assumes that the clumps are small compared to the Sobolev length.
	
	We performed simple tests to examine how the photon escape probability depends on the clump size when the volume-filling factor is held constant.  These tests indicate that the VFF approach is only applicable when the clump size is of order 0.2 Doppler widths, or less. This is true for both thick and thin lines. For thick lines, a clump size of less than a Doppler width allows significant escape of photons in the line wings, and in such cases the escape probability approaches the VFF value. 
	
	These results, combined with the remarkable success of the VFF approach in modelling WR stars, strongly suggest that clumps in WR winds are  highly fragmented, and that the difference between their lateral and radial sizes is relatively small (i.e. they are not pancakes). It is possible that some differences in shape may explain why the \ion{He}{II} profiles in some WR stars  are more rounded than predicted by models. Earlier studies have also found a need for the clumping structure to be broken on small angular scales. Assuming optically thin emission, \cite{Dessart2002_clump} and \cite{Dessart2002_3D} found, from an analysis of line profile variability, that clumps should have an angular size of about 3$^{\scriptstyle \rm o}$. Our simulations of line profiles, that allow for optical depth effects, suggest that the angular scale must be smaller than 3$^{\scriptstyle \rm o}$ -- the broken Shell model discussed earlier in the paper had an angular scale of 0.3$^{\scriptstyle \rm o}$ which yielded a lateral size for the clump similar to the radial size. More recent theoretical studies also suggest a small lateral scale although it is sensitive to the assumptions made in the simulations \citep{Sundqvist2018}. The same authors also find that the velocity dispersion in the lateral direction is much less than in the radial direction.
		
	Our spherical Shell models yield double-peaked profiles for many of the \ion{He}{II} profiles. While our models are admittedly contrived, they demonstrate that one cannot immediately assume that a double-peaked profile indicates a disk-like geometry. In the Shell models, the central absorption dip is a consequence of optical depth effects. \cite{Hillier2012} also demonstrated that the central reversals can be seen for some lines in rapidly rotating stars when non~LTE effects are considered.
	
	Anisotropic absorption effects arising from clumping can also affect X-ray line profiles \citep{Feldmeier2003,Oskinova2004,Sundqvist2012_por}, but are distinct from the process discussed here. In a smooth wind the red side of an X-ray line profile is generally more heavily absorbed than the blue side, simply because the photons have to transverse a much longer path length through the wind. However, the location of the emitting region, clumping and/or the presence of shells within the wind will then modify the asymmetry of the observed X-ray line profile. The effect is very much related to the global optical depth seen by an emitted photon. The effect discussed here influences photon escape from the resonance zone, and it is more localized than that associated with continuum processes influencing X-ray line profiles.
	
	Although our study provides insights into the ionization structure and spectral characteristics of HD~50896's clumpy wind, it  highlights a need to further investigate clumping and inhomogenities in stellar winds. Both 3D radiation-hydrodynamical simulations and 3D radiative transfer simulations are needed. Clumps (and the associated velocity field) are inherently a 3D phenomenon. While we were able to compute a spectrum for a multi-D clumped model, our Shell model is fundamentally a 1D model. Further, we make very simple assumptions about the velocity structure. In particular, we assume a monotonic velocity law and adopt a relatively large value for the microturbulence. These quantitatively affect our results. A simple illustration is provided by the \civdoubopt\ doublet which is pumped by the UV continuum at $\sim$312\,\AA. The presence of significant vorosity would weaken this line.
	
	In the present study we have generally ignored  the influence of the interclump medium, since its low density means that it has little influence on \ion{He}{ii} and \ion{He}{i}\ emission lines which are formed by recombination processes. In principle vorosity effects could weaken the absorption component on P~Cygni profiles (and the associated scattered component), as found in our modelling of AzV83. However in the dense wind modelled here, with its smooth velocity field, vorosity effects are reduced since the interclump medium may still be optically thick in certain transitions such as the \ion{N}{v} and \ion{C}{iv} resonance doublets.
	
    Recently a first attempt at 3D radiation-hydrodynamic simulations of Wolf-Rayet winds has been made by \cite{2022arXiv220301108M}.  Such simulations are providing insights into both the velocity law and clumping. However, the simulations of  \cite{2022arXiv220301108M} predict a volume-filling factor of around 0.5 which would seem to be too high compared with values required from spectral fitting.  This will need to be confirmed using detailed 3D spectral computations.

	\section*{Acknowledgements}
	Support from STScI theory grant HST-AR-14568.001-A (BLF, DJH) and  HST-AR-16131.001-A (DJH) is acknowledged. STScI is operated by the Association of Universities for Research in Astronomy, Inc., under NASA contract NAS 5-26555. We would especially like to thank Joachim Puls, and the referee, Jon Sundqvist, for comments which helped to improve the manuscript.

	\section*{Data availability}
    An earlier version of \cmfgen, and the associated atomic data, is available at \url{http://www.pitt.edu/~hillier}. Updates of this website are routinely made. Models underlying this article will be shared on reasonable request to the corresponding author.

	

	\bibliographystyle{mnras}
	\bibliography{shells_bib_HD50896}

	
	
	\appendix
	
	\section{Additional Tables}
	\label{appendix:A1}
	
	The models of HD~50896 were run using a small set of model atoms and abundances seen in Tables \ref{table:atom} and \ref{table:abund}. The atomic data used in the calculations, as well as atomic data sets for other species, are available for download from \url{http:/www.pitt.edu/~hillier}, and are routinely updated. In Table \ref{table:atom}, N$_F$ refers to the total number levels included in the atom and N$_S$ refers to the number of superlevels used in each atom. 
	
	Table \ref{table:abund} lists the abundances of the species found in Table \ref{table:atom}, where N(A) and X(A) denotes the relative number fraction and mass fraction of species A, respectively. The sources for the atomic data used in this paper are found in Appendix A of\, \cite{Flores2021}.
	
	\begin{table}
		\centering
		\caption{Summary of model atoms for several species.}
		\label{table:atom}
		\begin{tabularx}{\columnwidth}{*{5}{X}} 
			\hline
			&\textbf{Species} & \textbf{N$_S$}  &\textbf{N$_F$} &
			\\ \hline 
			
			&\ion{He}{I}      &  27  &  39 &    \\
			&\ion{He}{II}     &  13  &  30 &    \\
			&\ion{C}{IV}      &  14  &  14 &    \\
			&\ion{N}{III}     &  26  &  26 &    \\
			&\ion{N}{IV}      &  34  &  60 &    \\
			&\ion{N}{V}       &  45  &  67 &    \\
			&\ion{O}{IV}      &  10  &  10 &    \\
			&\ion{O}{V}       &  34  &  34 &    \\
			&\ion{O}{VI}      &  13  &  13 &    \\
			&\ion{Si}{IV}     &  22  &  33 &    \\
			&\ion{Fe}{IV}     &  21  & 280 &    \\
			&\ion{Fe}{V}      &  19  & 182 &    \\
			&\ion{Fe}{VI}     &  10  &  80 &    \\
			&\ion{Fe}{VII}    &  14  & 153 &    \\ \hline
		\end{tabularx}
	\end{table}
	
	\begin{table}
		\centering
		\caption{Summary of abundances of several species.}
		\label{table:abund}
		\begin{tabularx}{\columnwidth}{ *{4}{X} }
			\hline
			&\textbf{Abundance} &  &		\\ 
			\hline \\[-4pt]
			
			&$\frac{N(\text{C})}{N(\text{He})}$ & $1.0\times 10^{-4}$ &\\[5pt]

			&$\frac{N(\text{N})}{N(\text{He})}$ & $4.0\times 10^{-3}$ &\\[5pt]
			
			&$\frac{N(\text{O})}{N(\text{He})}$ & $1.0\times 10^{-4}$ &\\[5pt]
			
			&$\frac{N(\text{Si})}{N(\text{He})}$ & $1.3\times 10^{-4}$ &\\[5pt]
			
			&$\frac{N(\text{Fe})}{N(\text{He})}$ & $1.1\times 10^{-4}$ &\\[5pt]
			
			\hline\\
		\end{tabularx}
	\end{table}

\section{Constraint on the interclump medium}

\centerline{\Large Filling factor}
\label{app_f_constaint}

To place a constraint on the density of the interclump medium we assume that our Shell model must have the same effect on density, and density squared processes, as in the classic VFF approach. That is, the radial integral of the
density squared is preserved, while the radial integral of the density is scaled by the $\sqrt{f}$.  This will approximately preserve the equivalent width of (most) emission lines, but will scale the strength of the electron scattering wings by the $\sqrt{f}$. 

\blankhalf
Let: 
\mylist
$\rhosm$ be the density of the smooth unclumped medium that matches the  line  strengths of recombination lines (i.e. those having a density squared dependence), 
\mylist
$f$ be the classic volume-filling factor, 
\mylist
 $w$ be the width of the clump, 
\mylist
$\alpha$ be the contrast between the clump density and the interclump medium,
\noindent
and
\mylist
$\gamma$ = clump density / $\rhosm$. \\

\noindent
For the clumped medium to
match the result for the VFF approach we require for processes linearly
dependent on the density that:
\begin{equation}
\sqrt{f} \rhosm =   w. \gamma \rhosm + (1-w) \gamma \rhosm \alpha
\end{equation}
\noindent
or
\begin{equation}
\sqrt{f} =   w. \gamma + (1-w) \gamma  \alpha \,\,.
\label{eq_den}
\end{equation}

\noindent
For density squared processes  we require
\begin{equation}
\rhosm^2 =   w. \gamma^2 \rhosm^2 + (1-w) \gamma^2 \rhosm^2 \alpha^2
\end{equation}
or
\begin{equation}
1=  w. \gamma^2 +  (1-w) \gamma^2 \alpha^2 \,\,.
\label{eq_densq}
\end{equation}
Solving equations~\ref{eq_den} and \ref{eq_densq}
yields
\begin{equation}
f \ge {4\alpha \over (1+\alpha)^2}
\end{equation}
\noindent
which for small $f$ (large $\alpha$) can be rewritten as $\alpha > 4/f$.

	\section{Line Photon Escape Probability in a Slab}
	\label{app_esc_prob}
	
	In this section we derive the line optical depth and escape 
	probabilities for a photon from a slab (or shell) of finite width. For simplicity we only consider escape in a single direction. We further assume the slab that the opacity and velocity gradient are constant within the slab.
	
	The line optical depth for a photon of frequency $\nu$ in a finite slab of length $Z_{max}$ is given by 
	\begin{equation}
		\tau_{\nu}(Z) = \int_{Z=0}^{Z=Z_{max}} \chi(Z')\,\phi(\nu) \ \,dZ'
		\label{eq:tau}
	\end{equation}
	where $\chi(Z')$ is the opacity of the slab; in the case of a homogeneous slab, we can assume the opacity to be constant ($\chi(Z') = \bar{\chi} $). $\phi(\nu)$ is the absorption profile function given by 
	\begin{equation}
		\phi(\nu)\,d\nu =  \dfrac{1}{\sqrt{\pi}} e^{- u^2} \,du
	\end{equation}
	where, for convenience, we let $u(\nu) = (\nu - \nu_o)/\Delta\nu$ with $\Delta\nu = \varv_d \,\nu_o/c$ representing the Doppler width of the profile and $\nu_o$ is the rest frame frequency. In the presence of a velocity field, the photon's frequency is Doppler shifted, $\nu' = \nu(1 - \varv(Z)/c)$, giving us a new exponential term in the absorption profile, $u'(\nu) = (\nu' - \nu_o)/\Delta\nu$. Using $u$ and the definition of the Doppler width ($\Delta\nu = \varv_d \,\nu_o/c$), we rewrite $u'(\nu)$ to be a function of $u$ and $Z$ as follows
	\begin{equation}
		\begin{split}
		  u' = \dfrac{\nu' - \nu_o}{\Delta\nu} & = \dfrac{ \nu - \nu_o}{\Delta\nu} - \dfrac{\nu \, \varv(Z)}{c\,\Delta\nu} \\
		  & = u - \dfrac{d\varv}{dZ} \dfrac{Z}{c\,\Delta\nu} \, \big(u\Delta \nu + \nu_o\big) \\
		  & = u - \dfrac{d\varv}{dZ} \dfrac{Z}{\varv_d} \, \left( u\, \dfrac{\varv_d}{c} + 1 \right) \\
		  & \approxeq u - \dfrac{d\varv}{dZ} \dfrac{Z}{\varv_d} 
		\end{split}
		\label{eq:exp_u_prime}
	\end{equation}
	where, for the last line, we assumed $\varv_d /c << 1$. 
	
	To calculate the line optical depth, it is advantageous to perform the integration over $u'$ instead of $Z'$ as it makes the upcoming integral easier. Thus: 
	\begin{equation}
		\begin{split}
		  \tau(\nu) & = \int_{Z}^{Z_{max}} \dfrac{\chi}{\Delta\nu \sqrt{\pi}} \, \mathrm{exp}\left(- \big(\dfrac{\nu' - \nu_o}{\Delta\nu} \big)^2 \right) \ \,dZ' \\\\
		  & = \int_{u'(Z=Z_o)}^{u'(Z=Z_{max})} \dfrac{\bar{\chi}}{\Delta\nu \sqrt{\pi}} \, e^{- u^2} \ \, \left( - \varv_d \dfrac{dZ}{d\varv} du' \right) \\\\
		  & = \bar{\chi} \dfrac{\varv_d}{\Delta\nu} \left( \dfrac{dZ}{d\varv} \right)  \int_{u'(Z=Z_{max})}^{u'(Z=Z_o)} \dfrac{1}{\sqrt{\pi}} e^{- u^2} \, du' \\\\
		  {\rm and\,\,hence} \\
		  \tau(u,Z_o) & = \dfrac{\tau_o}{2} \left[ \, \mathrm{Erf}(\, u\,  ) - \mathrm{Erf}\left(\, u - \dfrac{d\bar{\varv}}{dZ} (Z_{max}-Z_o)\, \right)\,  \right] \, . \\
		\end{split}
		\label{eq:tau_2}
	\end{equation}
	The term in front of the integral is simply the Sobolev optical
	depth which we denote by $\tau_o$. We have also defined
	 $\bar{\varv} = \varv / \varv_d$. Since we are assuming a constant velocity gradient, $\dfrac{d\bar{\varv}}{dZ} (Z_{max}-Z_o)$ is simply the distance to the boundary in
	 units of the Doppler width.
	 Finally, the escape probability from $Z_o$ in the $Z$ direction, $\beta(Z_o)$, is given by
	\begin{equation}
		  \beta(Z_o) = \int_{-\infty}^{\infty} \phi(u) \, e^{-\tau(u,Z_o)}  \ \,du \, . 
		\label{eq:esc_prob}
	\end{equation}
	
	Finally, we take the Doppler-velocity average of the escape probability 	\begin{equation}
		  \left< \beta \right> =\dfrac{1}{V_d} \int_{0}^{V_d} \beta(z_o) \,d\bar{\varv}
		\label{eq:av_esc_prob}
	\end{equation}
	where $V_d$ is the total Doppler width that the slab spans.

	\bsp	
	\label{lastpage}
	
\end{document}